\documentclass[10pt,a4paper]{article}
\usepackage[utf8]{inputenc}
\usepackage[english]{babel}
\usepackage{amsmath}
\usepackage{comment}
\usepackage{amsfonts}
\usepackage{amssymb}
\usepackage{makeidx}
\usepackage{graphicx}
\usepackage{lmodern}
\usepackage{kpfonts}
\usepackage{bbm}
\usepackage{bm}
\usepackage{import}
\usepackage{appendix}
\usepackage{array}
\usepackage{caption}
\usepackage{subcaption}
\usepackage{lscape} 
\usepackage{enumerate}
\usepackage[dvipsnames]{xcolor}
\usepackage[left=3cm,right=3cm,top=2cm,bottom=2cm]{geometry}
\newcommand{\app}{``}

\newtheorem{theorem}{Theorem}[section]

\newtheorem{definition}[theorem]{Definition}

\newtheorem{remark}[theorem]{Remark}

\newtheorem{example}[theorem]{Example}

\newcommand{\be}{\begin{equation}}
\newcommand{\ee}{\end{equation}}

\begin{document}

\begingroup\parindent0pt
\centering
\begingroup\LARGE
\bf
Uphill in reaction-diffusion multi-species\\ interacting particles systems
\par\endgroup
 \vspace{3.5em}
 \begingroup\large
 {\bf Francesco Casini}, 
 {\bf Cristian Giardin\`{a}},
 {\bf Cecilia Vernia}
 \par\endgroup
\vspace{2em}
% \begingroup\sffamily\footnotesize
% 
% $^a\,$Laboratoire de Physique de l’École Normale Supérieure, \\CNRS, Université PSL, Sorbonne Universités, \\24 rue Lhomond, 75005 Paris, France\\
% % frassek@ihes.fr\\
% \vspace{1em}
% $^b\,$
University of Modena and Reggio Emilia, FIM,\\
 Via G. Campi 213/b, 41125 Modena, Italy\\
% \vspace{1em} %Université PSL, CNRS, Sorbonne Université,\\ Université Paris-Diderot, %Sorbonne Paris Cité, Paris, France\\
% \par\endgroup
\vspace{5em}

Version: \today

 \vspace{1.cm}
 \begin{abstract}
 \noindent
 We study reaction-diffusion processes with multi-species of particles and
 hard-core interaction. We add boundary driving to the system  by means of
 external reservoirs which inject and remove particles, thus creating 
 stationary currents. We consider the 
 condition that the time evolution of 
 the average occupation evolves as
 the discretized version of a system of coupled diffusive equations with 
 linear reactions. 
 In particular, we identify a specific one-parameter family of such linear
 reaction-diffusion systems where the hydrodynamic limit behaviour
 can be obtained by means of a dual process.
 We show that partial uphill diffusion is possible
 for the discrete particle systems on the lattice,
 whereas it is lost in the hydrodynamic limit.
 \end{abstract}

\endgroup

\thispagestyle{empty}
\setcounter{tocdepth}{2}

% \newpage
%\tableofcontents
% \newpage

\newpage
\section{Introduction}	
		
	\subsection{Motivation and description of results}
The aim of this paper is to study `uphill diffusion' in multi-species
interacting particle systems with hard-core interaction.
We analyse systems consisting of $n$ types of particles and add boundary
reservoirs injecting and removing particles. Here, uphill diffusion means
that mass flows from regions  with lower density to regions with higher
density. Uphill diffusion is thus a violation of Fick's law. This
phenomenon has been reported in a single-species system in the presence 
of a phase transition (see \cite{de2011fourier,colangeli2016latent,colangeli2017particle,colangeli2017microscopic,colangeliGibertiVernia} for 1D particle systems with Kac potentials and \cite{colangeliVerniaGibertiGiardina}
for 2D lattice gases related to the Ising model). In multicomponent
systems, uphill diffusion arises as a result of the competition between 
the  gradients of each species \cite{krishna,olandesi}.
The phenomenon whereby current in a stationary system is in a direction
opposite to an external driving field has also been named `absolute
negative mobility' in \cite{cividini2018driven}.
Multi-species particle systems have been much studied in the recent literature, especially in relation to the notion of duality \cite{franceschini2022orthogonal,zhou,belitsky2015self,belitsky2015quantum,belitsky2018self,borodin2022shift,kuanmulti,kuan2018algebraic,kuniba2016stochastic}.

 \smallskip

	For diffusive models with a partial uphill, transport of mass on a finite volume (here assumed to be the unit $d$-dimensional cube)	
	is often described by the continuity equation
	\begin{equation}
		\label{continuity}
		\frac{\partial}{\partial t}\rho = - \nabla \cdot J
	\end{equation}
	and the Fick's law
	\begin{equation}
		\label{fick}
		J = - \sigma \nabla \rho
	\end{equation}
	Here $\rho:[0,1]^d \times \mathbb{R}_{+} \to [0,1]$ is the density of mass,
	$J:[0,1]^d \times \mathbb{R}_{+} \to \mathbb{R}$ is the current,
	and $\sigma >0$ is the constant diffusivity coefficient.  Equations
	\eqref{continuity} and \eqref{fick} can be obtained
	as the hydrodynamical limit of diffusive
	interacting particle systems of ``gradient type'' \cite{presutti},
	such as the simple symmetric exclusion process or
	the Kipnis-Marchioro-Presutti model \cite{kipnis1982heat}.
	Fick's law \eqref{fick} tells us that the total flow is opposite to  the density gradient.
	
	For multi-component systems with $n$ species,
	considering the vectors ${\bm \rho} = (\rho^{(1)},\ldots,\rho^{(n)})$ and ${\bm J} = (J^{(1)},\ldots,J^{(n)})$, where $\rho^{(i)}(x,t)$
	and $J^{(i)}(x,t)$ denote the density and 
	the current of the i$^\text{th}$ species,
	the generalization of \eqref{continuity} and \eqref{fick} is
	\begin{equation}
		\label{n-continuity}
		\frac{\partial}{\partial t}{\bm \rho} = - \nabla \cdot {\bm J}
	\end{equation}
	and
	\begin{equation}
		\label{n-fick}
		{\bm J} = - {\bm \Sigma} \cdot \nabla {\bm \rho}.
	\end{equation}
%	Combining together (\ref{n-continuity}) and (\ref{n-fick})
%	one gets the system of linear partial differential equations
%	\begin{equation}\label{diffusive_Eqn}
%		\frac{\partial }{\partial t}{\bm \rho} = {\bm \Sigma} \Delta {\bm \rho}
%	\end{equation}
	where ${\bm \Sigma}$ is now the $n\times n$ matrix of diffusion and 'cross-diffusion' coefficients. When ${\bm \Sigma}$ is
	non-diagonal, then uphill diffusion is possible \cite{krishna}. We distinguish between the case
	of {\em `partial' uphill}, which is obtained
	when the current of one of the species 
	has the same sign of the gradient of that species, 
	and {\em `global' uphill},
	which arises when the total mass flows
	from a region of lower total density
	to a region of higher total density.
	
\medskip	
	
In this paper, we shall investigate partial uphill diffusion for hard-core multi-species interacting particle systems. 
Our analysis will have two targets: on one hand, we would like to understand conditions on the rates defining the microscopic dynamics so that the system is described by a linear reaction-diffusion structure on a regular lattice; on the other hand, we aim to understand if and how such
particle systems display partial uphill diffusion
in the large scale limit.
To achieve those targets we will consider the {\em average occupation} of each species, which is a proxy for the true density.
In the spirit of \cite{schutz} and \cite{wadati} we shall impose that the equations for the average occupation of the species are closed.
Furthermore, we shall require that the evolution of the average occupation is described by the a discretized version of \eqref{n-continuity} and \eqref{n-fick}. Actually, besides diffusion, we shall further include the
possibility of reaction terms, as described in the next subsection. Our main results can be summarized as follows:
\begin{itemize}
\item We  show that the request
of a linear reaction-diffusion structure on a regular lattice imposes constraints on the values of the ``diffusivity matrix'' and
the reaction coefficient (see Theorem \ref{THM_principal}). %As a consequence of this,
%there will not be global uphill diffusion, whereas 
%there might be partial uphill.

\item
We identify a specific multi-species interacting particle system (see again Theorem \ref{THM_principal}) for which 
the closure of correlation functions
is accompanied by duality  (see Section \ref{dualitySection}). To our knowledge, this is the first multi-species interacting particle system
with reaction {\em and} diffusion for which
one can prove the existence of a dual process
(see \cite{presutti} for a perturbative treatment
of reaction-diffusion in the presence of duality
for the sole diffusive dynamics).

\item 
Duality then leads to the proof
of the hydrodynamic limit with the standard
correlation functions method \cite{presutti}.
Surprisingly, we shall see that -- although the microscopic dynamics 
has non-zero `cross-diffusivity' terms -- macroscopically the empirical mass distribution of each species satisfies
 hydrodynamic PDE's where the species are coupled only by the reaction term. In other words, after a suitable space/time diffusive scaling, the diffusivity matrix is necessarily diagonal and therefore partial uphill is absent. 
This is consistent with \cite{kahane, gorban2011quasichemical} where it has been observed that the densities of eq. \eqref{n-continuity} and \eqref{n-fick} remain positive if and only if the cross diffusivity terms are null.

\end{itemize}

We conclude this introduction with a discussion 
about uphill diffusion for equations \eqref{n-continuity} and \eqref{n-fick} plus a linear reaction term.

%	Some comments are in order:
%	\begin{itemize}
%		\item[•] can we keep the linear structure of the equations by including (conservative!) linear reactions when the diffusivity matrix is non-diagonal?
%		\item[•] can we obtain coupled reaction-diffusion equations with global or partial uphill?
%		We restrict to studying the average densities (the hydrodynamic limit
%		for the empirical densities goes beyond the scope of this paper).
%	\end{itemize}

	\subsection{Steady state uphill diffusion in multi-component systems}\label{sec1.2}
	 %The equation that governs the linear diffusion of a species is given by (\ref{diffusive_Eqn}). This law, shows that the time variation of the concentration of a species, is driven by the action of the one dimensional laplacian, multiplied by a proper constant. By endowing this equation with Dirichlet boundary conditions (that physically represents the contact of the system with bath at fixed density) the following diffusion problem on the previous domain is obtained:
	%\begin{equation}\label{simpleDiffusion}
	%	\begin{cases}
		%	\frac{\partial }{\partial t} \rho =  \sigma_{11} \Delta \rho\\
		%	\rho(x,0)=\rho_{0}(x)\\
		%	\rho(0,t)=\rho_{L}\\
		%	\rho(1,t)=\rho_{R}
	%	\end{cases}
%	\end{equation}

Without loss of generality, we restrict ourselves to the case of two species
diffusing on the unit interval. Let us call $\rho^{(\alpha)}(x,t):
[0,1]\times [0,\infty)\rightarrow[0,1]$ the density of the  species 
$\alpha\in\{0,1,2\}$. We impose the constraint 
$\rho^{(0)}+\rho^{(1)}+\rho^{(2)}=1$, which will represent later the 
hard-core interaction of the associated interacting particle system. 
It is then enough to study the evolution of $\rho^{(1)}$ and $\rho^{(2)}$,
which will be assumed to be smooth functions.
We consider a Cauchy problem with Dirichlet boundary conditions, 
where each density is endowed with an initial datum $\rho^{(\alpha)}(x,0) = \rho^{(\alpha)}_{0}(x)$ and boundary conditions $\rho^{(\alpha)}(0,t) =\rho^{(\alpha)}_{L}$ and $\rho^{(\alpha)}(1,t)= \rho^{(\alpha)}_{R}$ for $ \alpha=1,2$.
We are interested in the stationary properties.
We consider 
	\begin{equation}\label{stronglyWeaklycoupledPDE}
			\begin{split}
	\partial_{t}\rho^{(1)}=\sigma_{11}\partial_{x}^2\rho^{(1)}+\sigma_{12}\partial_{x}^2\rho^{(2)}+\Upsilon\left(\rho^{(2)}-\rho^{(1)}\right)\\
	\partial_{t}\rho^{(2)}=\sigma_{21}\partial_{x}^2\rho^{(1)}+\sigma_{22}\partial_{x}^2\rho^{(2)}+\Upsilon\left(\rho^{(1)}-\rho^{(2)}\right)
			\end{split}
		\end{equation}
where $\Sigma$ is a constant positive
definite matrix
		\begin{equation}\label{DiffusionMatrix}
			\Sigma=\begin{pmatrix}
				\sigma_{11}&\sigma_{12}\\
				\sigma_{21}&\sigma_{22}
			\end{pmatrix}
		\end{equation}
The stationary diffusive currents are given by
	\begin{eqnarray}
		J^{(1)}(x)&=&-\sigma_{11}\partial_{x}\rho^{(1)}(x)-\sigma_{12}\partial_{x}\rho^{(2)}(x) \nonumber\\
		J^{(2)}(x)&=&-\sigma_{21}\partial_{x}\rho^{(1)}(x)-\sigma_{22}\partial_{x}\rho^{(2)}(x)
	\end{eqnarray} 
We distinguish  two cases:  
%	We can distinguish two sub-classes: 
	\begin{itemize}
		\item \textit{global uphill}: 
this happens when the boundary values of
the total boundary density $\rho_L = \rho_L^{(1)}+\rho_L^{(2)}$ and
$\rho_R = \rho_R^{(1)}+\rho_R^{(2)}$
and the total current $J(x) = J^{(1)}(x) + J^{(2)}(x)$
are such that either $\rho_L < \rho_R$ and
$J(x) >0$ $\forall x\in [0,1]$, or $\rho_L > \rho_R$ and
$J(x) <0$ $\forall x\in [0,1]$.
\item \textit{partial uphill for the i$^{\text{th}}$ species}: 
for boundary values $\rho_{L}^{(1)}, \rho_{L}^{(2)}, \rho_{R}^{(1)}, \rho_{R}^{(2)}\geq 0$, the system has stationary partial uphill diffusion for the species $i\in \{1,2\}$ if $\rho_{L}^{(i)}<\rho_{R}^{(i)}$ and $J^{(i)}(x)>0$ $\forall x\in [0,1]$, or if $\rho_{L}^{(i)}>\rho_{R}^{(i)}$ and $J^{(i)}(x)<0$ $\forall x\in [0,1]$.
\end{itemize}
Clearly, in the case where each density simply obeys a one dimensional heat equation
		\begin{equation}\label{totallyDecoupled}
			\begin{split}
	\partial_{t}\rho^{(1)}(x,t)=\sigma_{11}\partial_{x}^2\rho^{(1)}(x,t)\\	\partial_{t}\rho^{(2)}(x,t)=\sigma_{22}\partial_{x}^2\rho^{(2)}(x,t)
			\end{split}
		\end{equation}
 no uphill diffusion (neither global nor partial) is possible. 

Global uphill diffusion can be obtained by keeping the 
matrix $\Sigma$ diagonal and adding a reaction term, 
i.e.
		\begin{equation}\label{WeaklycoupledPDE}
			\begin{split}
\partial_{t}\rho^{(1)}=\sigma_{11}\partial_{x}^2\rho^{(1)}+\Upsilon\left(\rho^{(2)}-\rho^{(1)}\right)\\
	\partial_{t}\rho^{(2)}=\sigma_{22}\partial_{x}^2\rho^{(2)}+\Upsilon\left(\rho^{(1)}-\rho^{(2)}\right)
			\end{split}
		\end{equation}
This has been shown in \cite{olandesi} where 
the above equations have been obtained as the hydrodynamical limit of a switching interacting particle system, and the region with global uphill has been explicitly characterized. 

\begin{figure}[ht!]
	\centering
	\begin{subfigure}[b]{0.49\textwidth}
	\centering
	 \includegraphics[width=\textwidth]{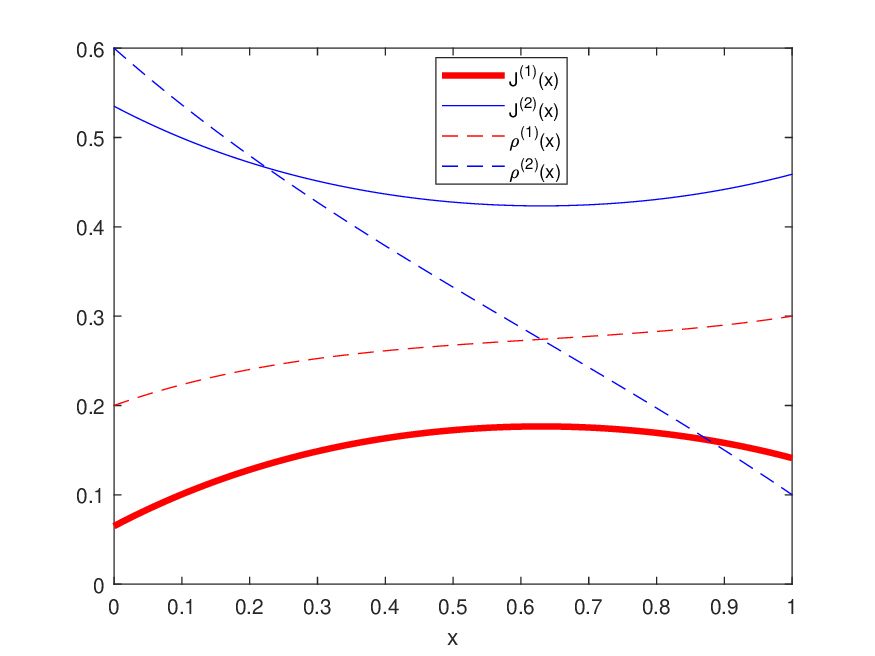}
	 
	% \label{caso1}
	\end{subfigure}
	 %\begin{subfigure}[b]{0.49\textwidth}
	%\centering
	% \includegraphics[width=\textwidth]{Region3.eps}
	% \caption*{ $(\rho_{L}^{(1)},\rho_{L}^{(2)},\rho_{R}^{(1)},\rho_{R}^{(2)})=(0.3,0.5,0.4,0.1)$}
	% \label{caso2}
	%\end{subfigure}
	 \caption{ Density profile (dashed lines) and currents (continuous line). The red color is for species 1 and the blue color for species 2. The boundary values are $(\rho_{L}^{(1)},\rho_{L}^{(2)},\rho_{R}^{(1)},\rho_{R}^{(2)})=(0.2,0.6,0.3,0.1)$. The diffusivity matrix and the
  reaction term are $\sigma_{11}=\sigma_{22}=\Upsilon=1$ and $\sigma_{12}=\sigma_{21}=1/2$. }
	 \label{fig:uno}
	 \end{figure}

To obtain partial uphill diffusion one needs to consider 
the more general case \eqref{stronglyWeaklycoupledPDE} with a {\em non-diagonal} matrix $\Sigma$. In Appendix \ref{sec7} we give the
stationary solution of \eqref{stronglyWeaklycoupledPDE}
from which the existence of partial uphill can be
ascertained.
Here we plot in Figure \ref{fig:uno} the stationary densities and currents for a specific choice of the boundary values and
of the diffusivity matrix and reaction term.
From the picture one can clearly see partial uphill diffusion
(in the absence of global uphill).
%For global uphill, let us observe that the total density %$\rho = \rho^{(1)}+\rho^{(2)}$ satisfies 
%	\begin{equation}\label{eqOlandesi}
%		\partial_{t}\rho=(\sigma_{11}+\sigma_{21})\partial^{2}_{x} \rho^{(1)}+(\sigma_{12}+\sigma_{22})\partial_{x}^{2}\rho^{(2)}
%	\end{equation}

%	{\color{blue}In this paper, the focus is on finding microscopic models such that the evolution equations for the {\em averaged} occupations are the discretized version of (\ref{stronglyWeaklycoupledPDE}), while discrretization of (\ref{totallyDecoupled}) and (\ref{WeaklycoupledPDE}) will be particular cases. \\
%	We will push further our analysis by studying the hydrodynamic limit and observing that, due to the diffusive rescaling, we always obtain weakly coupled limiting PDEs.}

	\subsection{Organization of the paper}

    Our paper is organized as follows.
	In Section \ref{sec2} we describe
	the generic form of a multi-species Markov process with constant rates allowing at most one particle per site.
	We define the process on a spatial structure given by a graph $G$
	and we compare to other models that have been studied in the literature. We then compute in Section \ref{sec3} the evolution equation for the average occupation
	variables of each species.
	
	From Section \ref{sec4}	onward we specialize to the case of {\em two species on one-dimensional chains}.
	We start, in Section \ref{sec4},  by imposing that
   the average occupations evolve as the discretized version of \eqref{stronglyWeaklycoupledPDE}.
	This leads to a linear algebraic system, which can be solved.
	As a result, sufficient and necessary conditions
    on the diffusivity matrix $\Sigma$
	and the reaction coefficient $\Upsilon$ 
    in order to have the discrete structure of a linear reaction-diffusion are identified
	in Theorem \ref{THM_principal}.
	Furthermore, it is shown in the same
	theorem an explicit example of a one-parameter family of symmetric processes having such linear and discrete reaction-diffusion structure.
	This specific model is further analyzed
	in Section \ref{dualitySection}, where we prove
	duality and the hydrodynamic limit. Section \ref{sec8} draws the conclusions
	of our analysis.

	\section{Hard-core multi-species particles on a graph $\mathbf{G=(V,E)}$}
	\label{sec2}
	
	{\bf Notation:} In what follows, we use greek letters
	($\alpha,\beta,\gamma,\delta, \ldots$)
	to denote the species of the particles and latin letters
	($x,y,z,\ldots$) to denote the sites of the graph.
	
	\bigskip
	
	In this section we  define our microscopic model on a generic graph $G=(V,E)$. Here, the set $V=\{1,2,\ldots,N\}$ is a collection of $N$ vertices. The set of edges $E$ is such that the graph is connected, directed and without self-edges.
	On this graph $G$ we consider a system of interacting particles, each of which has its own type/species. We assume there are $n$ species. Furthermore, on each vertex of the graph there is at most one particle (hard-core exclusion rule).
	Thus, the occupation variable at each vertex takes values
	in $\{0,1,2,\ldots n\}$, with
	type $0$ denoting the empty site.
	
\begin{figure}[ht]
    \centering
    \includegraphics[scale=0.5]{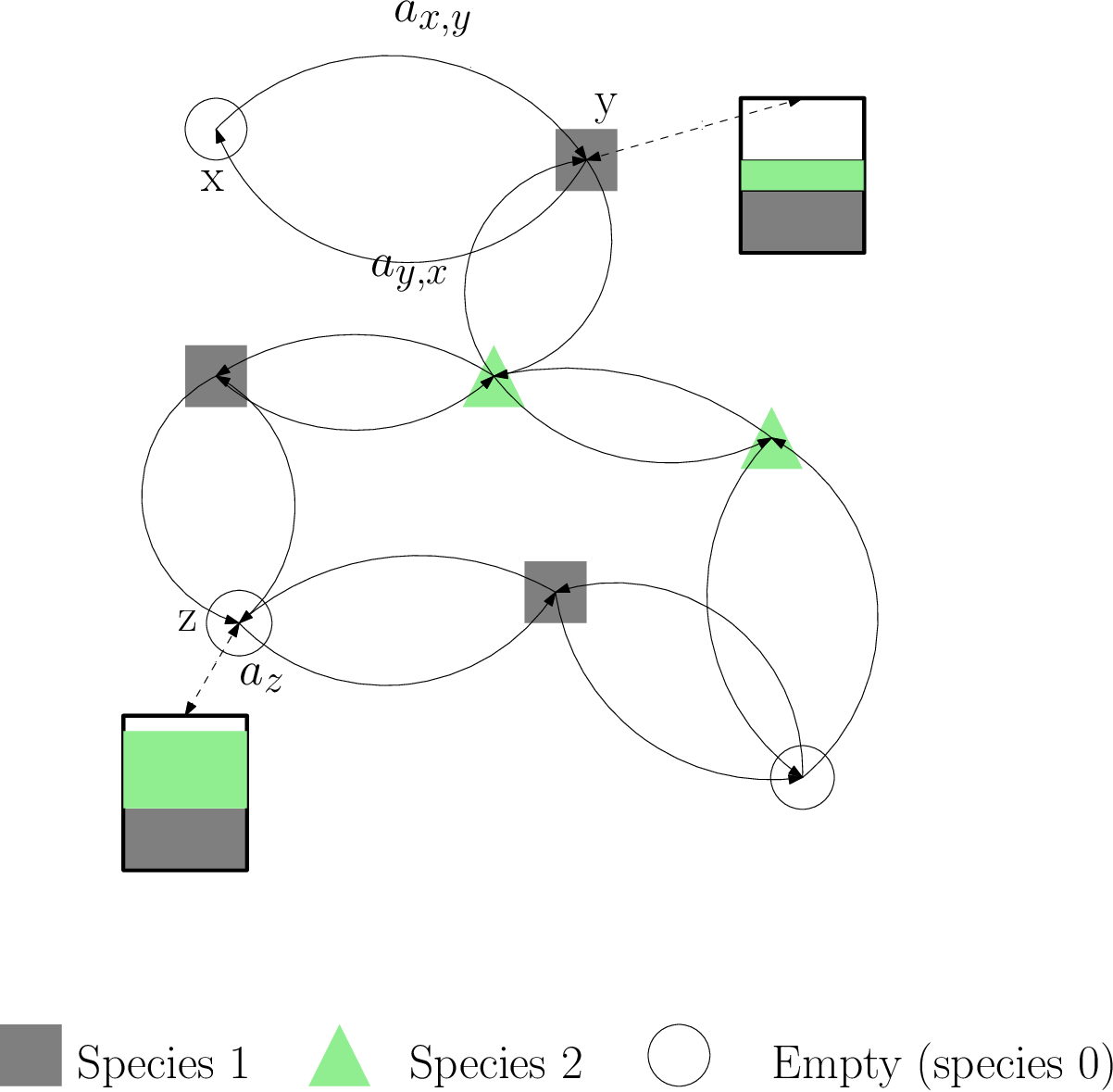}
    \caption{Hard-core two-species particles on a graph with 8 vertices and 2 reservoirs.
    Grey squares identify the species 1, green triangles the species 2, and white circles the empty state. The reservoirs are represented by rectangles, where the interior colours denote the density of species.}
    \label{fig:graph}
\end{figure}

	The dynamical rule is due to a one-body interaction and a two-body interaction:
	\begin{itemize}
	\item[-] 
	on each site $x\in V$ the occupation of type $\gamma$ changes to type $\alpha$ at rate  $a_x W_{\gamma}^{\alpha}(x)$; 
	\item[-]
	on each edge $(x,y)\in E$ the occupations of type $(\gamma,\delta)$ changes to type $(\alpha,\beta)$ at rate $a_{x,y} \Gamma_{\gamma\delta}^{\alpha\beta}$.
	\end{itemize}
		Here the non-negative numbers $\{a_{x,y}\}_{(x,y)\in E}$ and $\{a_{x}\}_{x\in V}$ are, respectively,
	edge weights (conductances) and site weights (local inhomogeneities) of the graph.
		For a visual representation of the process with two species see Figure \ref{fig:graph}.

	\subsection{Process definition}
	\label{sec2.1}

%	It is assumed that the process takes place on a finite dimensional (in general non planar), connected and undirected graph indicated by $G=(V,E)$, where $V=\{1,2,\ldots,N\}$ is the set of indices and $E$ is the set of edges between two indices. The phase space is given by $\{0,1,2,\ldots,n\}$. 
On the graph $G=(V,E)$, we consider the Markov process $\{\eta(t) ; t \ge 0\}$ 
with state space $\Omega=\{0,1,2,\ldots,n\}^{V}$.
A configuration of the process is denoted by $\eta=\left(\eta_{x}\right)_{x\in V}$, where each component can take the values $\eta_{x}\in \{0,1,...,n\}$ and where $\eta_{x}=\alpha$ means the presence of the species $\alpha$ at the site $x$. We recall that $\eta_x =0$ is interpreted as an empty site. 
The process is defined by the generator $\mathcal L$
working on functions $f: \Omega \to \mathbb{R}$ as
	\be 
	\label{gen-gen}(\mathcal{L}f)(\eta)
	=(\mathcal{L}_{edge}f)(\eta)+(\mathcal{L}_{site}f)(\eta),
	\ee
	where
	$$    (\mathcal{L}_{edge}f)(\eta)=\sum_{(x,y)\in E}a_{x,y}\cdot ({\mathcal{L}}_{x,y}f)(\eta) $$
	and
	$$(\mathcal{L}_{site}f)(\eta)= \sum_{x\in V}a_{x}\cdot (\mathcal{L}_{x}f)(\eta)
	$$
	We shall explain the two generators
	$\mathcal{L}_{edge}$ and $\mathcal{L}_{site}$
	in the following subsections.

	\subsubsection{The edge generator}
%	For the sake of simplicity we shall restrict to an edge generator which is the sum (weighted by the conductances) of homogeneous  terms.
	We introduce the $(n+1)^2\times (n+1)^2$ matrix $\Gamma$ whose elements are
	rates of transition for the particle jumps on each edge. More precisely, we denote by 
	$\Gamma_{\gamma\delta}^{\alpha\beta}$ the rate
	to change the configuration $\eta$ with
	$\eta_x=\gamma,\eta_y=\delta$ to the
	configuration $\eta'$ with
	$\eta'_x=\alpha,\eta'_y=\beta$, while $\eta'_z = \eta_z$
	for all $z\neq x,y$.
%	\begin{equation*}
%		(\eta_{1},\ldots,\underbrace{\gamma}_{x},\ldots,\underbrace{\delta}_{y},\ldots,\eta_{N}) \;{\longrightarrow}\; (\eta_{1},\ldots,\underbrace{\alpha}_{x},\ldots,\underbrace{\beta}_{y},\ldots,\eta_{N}) \qquad\qquad \text{at rate }\;\Gamma_{\gamma\delta}^{\alpha\beta}
%	\end{equation*} 
Thus, the single-edge generator is given by 
	\begin{eqnarray}
		\label{generatorBulk}
		&&\mathcal{L}_{x,y}f(\eta_{1},\ldots,\gamma,\ldots,\delta,\ldots,\eta_{N })\nonumber\\
		&&
		=\sum_{\alpha,\beta=0}^{n}\Gamma_{\gamma\delta}^{\alpha\beta}\left[f(\eta_{1},\ldots,\alpha
		%\underbrace{\alpha}_{x}
		,\ldots,\beta
		%\underbrace{\beta}_{y}
		,\ldots,\eta_{N})-f (\eta_{1},\ldots,\gamma
		%\underbrace{\gamma}_{x}
		,\ldots,\delta
		%\underbrace{\delta}_{y}
		,\ldots,\eta_{N})\right]
	\end{eqnarray}
	where  
	\begin{align*}
		\Gamma_{\gamma\delta}^{\alpha\beta}\geq 0\qquad &\text{if}\quad(\alpha,\beta)\neq (\gamma,\delta)\\
		\sum_{(\gamma,\delta)\in \{0,1,2,\ldots,n\}^{2}\;:\;(\gamma\delta)\neq (\alpha,\beta)}\Gamma_{\gamma\delta}^{\alpha\beta}=-\Gamma_{\alpha\beta}^{\alpha\beta}\quad&\forall(\alpha,\beta)\in \{0,1,2,\ldots,n\}^2\,.
	\end{align*}

	\subsubsection{The site generator}
	Having in mind that the site generator
	will describe a `boundary' driving 
	leading the system to a non-equilibrium steady
	state, we assume that on each site
	there is a process which injects and removes particles at a rate which is space-dependent. Thus, 
	for each vertex $x\in V$, we introduce the $(n+1)\times (n+1)$ matrix $W(x)$ whose elements are
	rates of transitions on that vertex. More precisely, we denote by 
	$W_{\gamma}^{\alpha}(x)$ the rate
	to change the configuration $\eta$ with
	$\eta_x=\gamma$ into the
	configuration $\eta'$ with
	$\eta'_x=\alpha$, while $\eta'_z = \eta_z$
	for all $z\neq x$.
%	, i.e.
%	\begin{equation*}
%		(\eta_{1},\ldots,\underbrace{\gamma}_{x},\ldots,\eta_{N}) \;{\longrightarrow}\; (\eta_{1},\ldots,\underbrace{\alpha}_{x},\ldots,\eta_{N}) \qquad\qquad \text{at rate }\;W_{\gamma}^{\alpha}(x)
%	\end{equation*} 
The single-vertex generator is given by 
	\begin{eqnarray}
		\mathcal{L}_{x}f(\eta_{1},\ldots,\gamma,\ldots,\eta_{N})=&\sum_{\alpha=0}^{n}W_{\gamma}^{\alpha}(x)\left[f(\eta_{1},\ldots,\alpha
		%\underbrace{\alpha}_{x}
		,\ldots.,\eta_{N})-f(\eta_{1},\ldots,
		\gamma
		%\underbrace{\gamma}_x
		,\ldots,\eta_{N})\right]
	\end{eqnarray}
	where 
	\begin{align*}
		W_{\gamma}^{\alpha}(x)\geq 0\quad &\text{if}\quad\alpha\neq \gamma\\
		\sum_{\gamma\in \{0,1,2,\ldots,n\}: \gamma\neq \alpha}W_{\gamma}^{\alpha}(x)=-W_{\alpha}^{\alpha}(x)\quad&\forall\alpha\in \{0,1,2,\ldots,n\}\,.
	\end{align*}

	\subsection{Comparison to other processes}
	Here, we discuss the relation of the general dynamics described above to some multi-species processes considered in the past literature (we consider here the case of homogeneous conductances and inhomogeneities $a_{x,y} = a_x = 1$).
	We shall mostly limit the discussion to {\em symmetric}
	systems (for asymmetric models there is also a large literature, see for instance \cite{franceschini2022orthogonal} and references therein). In most cases, previous analyses have been restricted to a regular lattice or a one-dimensional chain. 
		\begin{itemize}
		
		\item \textit{General multi-species models}. The  edge dynamics of the  reaction-diffusion particle system in Section \ref{sec2.1}  has been considered 
		on a d-dimensional lattice in   \cite{schutz} for the case $n=1$ species
		and in \cite{wadati}
		for the case of an arbitrary number of species.
		In  those papers, sufficient conditions on the rates $\Gamma_{\gamma\delta}^{\alpha\beta}$ to guarantee the existence of dual process have been identified.
		
		\item \textit{Multi-species exclusion processes}. The edge dynamics of multi-species    simple symmetric exclusion processes on a d-dimensional lattice, 
		with at most one-particle per site, has been considered in
	\cite{quastel}. It corresponds
		to the model of Section \ref{sec2.1} with
	$\Gamma_{0\alpha}^{\alpha 0} = \Gamma_{\alpha 0}^{0\alpha}\neq 0$ for all  $\alpha=0,1,\ldots,n$,
	while all other off-diagonal elements of the matrix $\Gamma$ vanish, as well as the elements of the matrices $W(x)$.	For this model, the hierarchy of equations
	for the correlations does not close, and the hydrodynamic limit has been shown in \cite{quastel} to be given
	by two coupled {\em non-linear} heat equations. An open boundary version of the model with simple symmetric exclusion dynamic in the bulk has been presented in \cite{brzank2006boundary}. It corresponds to the model of Section 2.1 with $\Gamma_{b 0}^{0 b}=\Gamma_{0 b}^{b 0}=D_{b}$ and with boundary rates $W^{b}_{0}(1)=\alpha_{b}, W_{b}^{0}=\gamma_{b}, W_{0}^{b}(N)=\beta_{b}, W_{b}^{0}(N)=\delta_{b}$ (here $b$ labels the species). All the other off-diagonal elements $\Gamma$ and $W(z)$ vanish.

		\item \textit{Multi-species stirring process}.
		In the stirring process \cite{arita2009spectrum,arndt1998spontaneous}, every couple of types
		is exchanged  in position with the same rate,
		which can be taken equal to 1 without loss of generality.
		Thus, the bulk dynamics of the stirring process corresponds
		to the case		$\Gamma_{\gamma\delta}^{\delta\gamma}=1$ for all $\gamma,\delta=0,1,\ldots,n$,
	while all other off-diagonal elements of the matrix $\Gamma$ vanish.
	The hydrodynamic limit
	of the stirring process on a lattice is given by
	$n$ independent diffusions, i.e.
	the generalization of \eqref{totallyDecoupled} to $n$ types. The multi-species stirring process on a chain with boundary driving has been studied in \cite{vanicat}
	with the choice $W_{\gamma}^{b}(1) = \alpha_b$
	and $W_{\gamma}^{b}(N) = \beta_b$.
	With this particular choice of the boundary rates the model is solvable and correlation functions in the non equilibrium steady state have been computed using the matrix product ansatz.  
		\end{itemize}

%	 It is interesting to compare this dynamics with some others that have been studied in the past. In the following, a quick (and non exhaustive) comparison with the bulk of some models (for simplicity by considering the case on the line graph) that have a similar dynamics in some sense:
	\begin{itemize}
%		\item \textit{Multispecies simple exclusion dynamics}: this dynamics has also been studied in \cite{brzank2006boundary} and in \cite{quastel}. It consists, in an hard core exclusion between the two species. Species $A$ and $B$ of particles can hop on their nearest neighbors only if they are empty. At most one particle per site is allowed. Thus, by comparing with the above process, just the rates $\Gamma_{0\alpha}^{\alpha 0}$ and $\Gamma_{\alpha 0}^{0\alpha}$ $\forall \alpha=0,1,\ldots,n$ are present.
%		\item \textit{Stirring process}: in references \cite{arita2009spectrum}, \cite{arndt1998spontaneous} and \cite{vanicat} the bulk dynamics is based on exchanges between nearest neighbor sites. While in the last reference, all the exchanges are allowed, in the second one just some of them are possible. Finally, in the first one, a non symmetric exchange is assumed. These models, are particular cases of the one described above when the rates are just $\Gamma_{\gamma\delta}^{\delta\gamma}$ $\forall \gamma,\delta=0,1,\ldots,n$.
%		\item \textit{partial uphill}: in the reference \cite{schutz}, the model is the same of the one studied in \cite{wadati}, but with the case of just one species performing the dynamics, thus their rates are $\Gamma_{\gamma\delta}^{\alpha\beta}$ but with $\alpha,\beta,\gamma,\delta=0,1$
		\item \textit{Multi-species switching process}: A different set-up 
		for multi-species particle systems has been recently proposed in \cite{olandesi,redig2022ergodic}. One considers $n$ ``piled'' copies of the graph $G$, each with its own single-type dynamics. The possibility of changing type is described by a \textit{switching rate} between layers. This set-up eliminates the constraint of
		one particle per site, in the sense that the projection of the dynamics on the columns of the piled graph allows
		the presence of several particle of different types on the same ``base'' site. In the case
        where each layer is
        a one-dimensional
        chain and two-layers are considered, the hydrodynamic limit
        has been shown to be given by the ``weakly'' coupled reaction diffusion equation   \eqref{WeaklycoupledPDE}.
        When boundary reservoirs are added, global uphill diffusion and boundary layers are possible \cite{olandesi}.

%		From a visual point of view, one can say that the process is defined on two different lines (one per particle species). On each line the dynamics is the exclusion one (with possible different rates), while there is the possibility for a particle of changing line with a proper switching rate.
%		The state space is given by 
%		\begin{equation*}
%			\Omega=\{0,1\}^{N}\times\{0,1\}^{N}
%		\end{equation*}
%		For these reasons, the possibility of having on the same site two different species of particles is left open. \\
%		In the case described in this paper, the dynamics is substantially different. Indeed, the two species of particles work on a common line and on each site at most one particle (of any species) is allowed. Thus it is not possible having two different particles at the same time in a fixed site. This difference will have an effect on the study of the uphill. 
	\end{itemize}
	\section{Evolution equations for the average occupation}
	\label{sec3}
	For the model introduced in Section \ref{sec2.1}, we define the average of the occupation variable of each species $\zeta \in \{0,1,\ldots,n\}$ 
	at time $t\ge 0$ and at the vertex $z\in V$
	\begin{equation}
	   \label{ave} \mu_{z}^{(\zeta)}(t)=\mathbb{E}\left[\mathbbm{1}_{\{\mathcal{I}^\zeta_z\}}(\eta(t))\right].
	\end{equation}
	Similarly, we consider the time-dependent correlations between
	species $\zeta,\zeta' \in \{0,1,\ldots,n\}$ 
	at points $z,z'\in V$
	\begin{equation}
	   \label{corr} c_{z,z'}^{(\zeta,\zeta')}(t)=\mathbb{E}\left[\mathbbm{1}_{\{\mathcal{I}^\zeta_z\}}(\eta(t))\mathbbm{1}_{\{\mathcal{I}^{\zeta'}_{z'}\}}(\eta(t))\right].
	\end{equation}
	Here $\mathcal{I}_{z}^{\zeta} = \{\eta\in\Omega \,:\;\eta_z=\zeta\}$ and $\mathbbm{1}_{\mathcal{I}}$ denotes the indicator function of the set $\mathcal{I}$. The notation $\mathbb{E} \left[ f(\eta(t))\right] = \int \nu_0(d \eta) \mathbb{E}_{\eta}\left[f(\eta(t))\right]$
	denotes the expectation in the process $\{\eta(t)\}_{t\ge 0}$ started from the initial measure $\nu_0$.
	The evolution equation of the density of the $\zeta$-species can be obtained by acting 
	with the generator. We have
	\begin{equation}
	\label{pippobaudo}
		\frac{d \mathbb{E}\left[\mathbbm{1}_{\{\mathcal{I}^\zeta_{z}\}}(\eta(t))\right]}{dt}=\mathbb{E}\left[\left(\mathcal{L}\mathbbm{1}_{\{\mathcal{I}^\zeta_{z}\}}\right)(\eta(t))\right].
	\end{equation}
	In the following section we evaluate the right hand side of this equation by considering first edge contributions and then site contributions.
	\subsection{Action of $\mathcal{L}_{x,y}$}
	
	If $z\notin \{x,y\}$ then obviously $\left(\mathcal{L}_{x,y}\mathbbm{1}_{\{\mathcal{I}_{z}^{\zeta}\}}\right)(\eta) = 0$. Otherwise, recalling that the graph $G$ is directed and the notation of \cite{wadati}, 
	we have the following: 
	when we fix $z=x$ then %\footnote{{\color{blue}All the terms with subscript "3" are not involved in the action of the generator on one site occupation variable, thus we do not consider them in this section}}
	\begin{equation}
	\label{eq:16}
		\left(\mathcal{L}_{z,y}\mathbbm{1}_{\{\mathcal{I}_{z}^{\zeta}\}}\right)(\eta)=A^{\zeta}_{1}+\sum_{\delta=1}^{n}F_{+1}^{\zeta\delta}\mathbbm{1}_{\{\mathcal{I}_{y}^{\delta}\}}(\eta)+\sum_{\gamma=1}^{n}B_{1}^{\zeta\gamma}\mathbbm{1}_{\{\mathcal{I}_{z}^{\gamma}\}}(\eta)+\sum_{\gamma,\delta=1}^{n}G_{+1}^{\zeta\gamma\delta}\mathbbm{1}_{\{\mathcal{I}_{y}^{\gamma}\}}(\eta)\mathbbm{1}_{\{\mathcal{I}_{z}^{\delta}\}}(\eta)
	\end{equation}
	and when we fix  $z=y$ then
	\begin{equation}
	\label{eq:17}
		\left(\mathcal{L}_{x,z}\mathbbm{1}_{\{\mathcal{I}_{z}^{\zeta}\}}\right)(\eta)=A^{\zeta}_{2}+\sum_{\gamma=1}^{n}F_{-1}^{\zeta\gamma}\mathbbm{1}_{\{\mathcal{I}_{x}^{\gamma}\}}(\eta)+\sum_{\delta=1}^{n}C_{2}^{\zeta\delta}\mathbbm{1}_{\{\mathcal{I}_{z}^{\delta}\}}(\eta)+\sum_{\gamma,\delta=1}^{n}G_{-1}^{\zeta\gamma\delta}\mathbbm{1}_{\{\mathcal{I}_{z}^{\gamma}\}}(\eta)\mathbbm{1}_{\{\mathcal{I}_{x}^{\delta}\}}(\eta) 
	\end{equation}
	where the constants are defined as follows: 
	\begin{enumerate}
		\item \textit{zero-order terms}:
		\begin{align*}
			A^{\zeta}_{1}&=\sum_{\beta=0}^{n}\Gamma_{00}^{\zeta\beta} \qquad\qquad
			A^{\zeta}_{2}=\sum_{\beta=0}^{n}\Gamma_{00}^{\beta\zeta}
		\end{align*}
		
		\item \textit{first-order terms}: 
		\begin{align*}
			B_{1}^{\zeta\gamma}&= \begin{cases}
				\sum_{\beta=0}^{n}(\Gamma_{\gamma 0}^{\zeta\beta}-\Gamma_{00}^{\zeta\beta})\;\;\;\;\quad\qquad\qquad\quad \text{if }\;\zeta\neq \gamma\\
				-\sum_{\beta=0}^{n}\left(\sum_{\zeta^{'}=0\;:\;\zeta^{'}\neq \zeta}^{n}\Gamma_{\zeta 0}^{\zeta^{'}\beta}+\Gamma_{00}^{\zeta\beta}\right)\;\;\;\;\;\text{if }\;\zeta= \gamma\\
			\end{cases}\\
			C_{2}^{\zeta\delta}&=\begin{cases}
				\sum_{\beta=0}^{n}(\Gamma^{\beta\zeta}_{0\delta}-\Gamma_{00}^{\beta\zeta})\qquad\qquad\qquad\;\;\;\;\;\text{if}\;\zeta\neq \delta\\
				-\sum_{\beta=0}^{n}\left(\sum_{\zeta^{'}=0\;:\;\zeta^{'}\neq \zeta}^{n}\Gamma_{0\zeta}^{\beta\zeta^{'}}+\Gamma_{00}^{\beta\zeta}\right)\;\;\;\;\;\text{if}\;\zeta= \delta\\
			\end{cases}\\
			F_{-1}^{\zeta\gamma}&=
			B_{2}^{\zeta\gamma}
			=\sum_{\beta=0}^{n}(\Gamma_{\gamma 0}^{\beta\zeta}-\Gamma_{00}^{\beta\zeta})\\
			F_{+1}^{\zeta\delta}&
			=C_{1}^{\zeta\delta}
			=\sum_{\beta=0}^{n}(\Gamma_{0\delta}^{\zeta\beta}-\Gamma_{00}^{\zeta\beta})
		\end{align*}
		\item \textit{second-order terms:}
		\begin{align*}
			G_{+1}^{\zeta\gamma\delta}&=D_{1}^{\zeta,\gamma,\delta}=\begin{cases}
				\sum_{\beta=0}^{n}(\Gamma_{\gamma\delta}^{\zeta\beta}-\Gamma_{\gamma 0}^{\zeta\beta}-\Gamma_{0\delta}^{\zeta\beta}+\Gamma_{00}^{\zeta\beta});\;\;\;\;\qquad\qquad\qquad\qquad\qquad\qquad\qquad\text{if }\;\zeta\neq \gamma\\
				-\sum_{\beta=0}^{n}\left(\sum_{\zeta^{'}=0\;:\;\zeta^{'}\neq \zeta}^{n}\Gamma_{\zeta\delta}^{\zeta^{'}\beta}+\Gamma_{0\delta}^{\zeta\beta}\right)+\sum_{\beta=0}^{n}\left(\sum_{\zeta^{'}=0\;:\;\zeta^{'}\neq \zeta}^{n}\Gamma_{\zeta 0}^{\zeta^{'}\beta}+\Gamma_{00}^{\zeta\beta}\right)\;\;\;\; \text{if }\;\zeta= \gamma
			\end{cases}
        \end{align*}
			%D_{2}^{\zeta\gamma\delta}=
			\begin{align*}
						G_{-1}^{\zeta\gamma\delta}&=D_{2}^{\zeta,\gamma,\delta}=
			\begin{cases}
				\sum_{\beta=0}^{n}(\Gamma_{\gamma\delta}^{\beta\zeta}-\Gamma_{\gamma 0}^{\beta\zeta}-\Gamma_{0\delta}^{\beta\zeta}+\Gamma_{00}^{\beta\zeta})\;\;\;\;\qquad\qquad\qquad\qquad\qquad\qquad\qquad\text{if }\;\zeta\neq \delta\\
				-\sum_{\beta=0}^{n}\left(\sum_{\zeta^{'}=0\;:\;\zeta^{'}\neq \zeta}^{n}\Gamma_{\gamma\zeta}^{\beta\zeta^{'}}+\Gamma_{\gamma 0}^{\beta\zeta}\right)+\sum_{\beta=0}^{n}\left(\sum_{\zeta^{'}=0\;:\;\zeta^{'}\neq \zeta}^{n}\Gamma_{0\zeta }^{\beta\zeta^{'}}+\Gamma_{00}^{\beta\zeta}\right)\;\;\;\;\;\text{if }\;\zeta= \delta
			\end{cases}
		\end{align*}
	\end{enumerate}
	\subsection{Action of $\mathcal{L}_{x}$}
	If $z\neq x$ then obviously $\left(\mathcal{L}_{x}\mathbbm{1}_{\{\mathcal{I}_{z}^{\zeta}\}}\right)(\eta) = 0$. Otherwise
	
	%\begin{equation}
	%	\left(\mathcal{L}_z\mathbbm{1}_{\{\mathcal{I}_{z}^{1}\}}\right)(\eta)
	%	=
	%	\overbrace{W_0^{1}(z)}^{A^{1}(z)} +\left(\overbrace{-W_0^{1}(z) - W_{1}^0(z)- W_{1}^2(z)}^{F^{11}(z)}\right)\mathbbm{1}_{\{\mathcal{I}_{z}^{1}\}}(\eta)+\left(\overbrace{W_2^{1}(z) -W_0^{1}(z)}^{F^{12}(z)}\right)\mathbbm{1}_{\{\mathcal{I}_{z}^{2}\}}(\eta)
	%\end{equation}
	%\begin{equation}
	%	\left(\mathcal{L}_z\mathbbm{1}_{\{\mathcal{I}_{z}^{2}\}}\right)(\eta)
	%	=
	%	\overbrace{W_0^{2}(z)}^{A^{2}(z)} +\left(\overbrace{-W_0^{2}(z) - W_{2}^0(z)- W_{2}^1(z)}^{F^{22}(z)}\right)\mathbbm{1}_{\{\mathcal{I}_{z}^{2}\}}(\eta)+\left(\overbrace{W_1^{2}(z) -W_0^{2}(z)}^{F^{21}(z)}\right)\mathbbm{1}_{\{\mathcal{I}_{z}^{1}\}}(\eta)
	%\end{equation}
	
		\begin{equation}
		\label{eq:18}
		\left(\mathcal{L}_{z}\mathbbm{1}_{\{\mathcal{I}_{z}^{\zeta}\}}\right)(\eta)=A^{\zeta}(z)+\sum_{\beta=1}^{n}F^{\zeta\beta}(z)\mathbbm{1}_{\{\mathcal{I}_{z}^{\beta}\}}(\eta)
	\end{equation}
where now the constants are defined as:
	\begin{enumerate}
		\item \textit{zero-order term}: 
		\begin{align*}
			A^{\zeta}(z)&=W_{0}^{\zeta}(z)
		\end{align*}  
		\item \textit{first-order term}:
		\begin{equation*}
		F^{\zeta\beta}(z)=\begin{cases}	W_{\beta}^{\zeta}(z)-W_{0}^{\zeta}(z)\qquad\qquad\qquad\qquad\;\;\;\text{if }\;\zeta\neq\beta\\	-\sum_{\zeta^{'}=0\;:\;\zeta^{'}\neq \zeta}^{n}W_{\zeta}^{\zeta^{'}}(z)-W_{0}^{\zeta}(z)\qquad\text{if }\;\zeta=\beta
				\end{cases}.
		\end{equation*}
	\end{enumerate} 

	\subsection{Action of $\mathcal{L}$}
	We now collect the results of the previous sections. 
	We may write
	\begin{eqnarray}
		\left(\mathcal{L}\mathbbm{1}_{\{\mathcal{I}_{z}^{\zeta}\}}\right)(\eta) 
		& = & 
		\sum_{x,y\;:\;(x,y)\in E} a_{x,y}\left(\mathcal{L}_{x,y}\mathbbm{1}_{\{\mathcal{I}_{z}^{\zeta}\}}\right)(\eta) 
		+
		\sum_{x} a_x \left(\mathcal{L}_x\mathbbm{1}_{\{\mathcal{I}_{z}^{\zeta}\}}\right)(\eta) \nonumber\\
		& = & 
		\sum_{y\;:\;(z,y)\in E} a_{z,y}\left(\mathcal{L}_{z,y}\mathbbm{1}_{\{\mathcal{I}_{z}^{\zeta}\}}\right)(\eta) 
		+
		\sum_{x\;:\; (x,z)\in E} a_{x,z}\left(\mathcal{L}_{x,z}\mathbbm{1}_{\{\mathcal{I}_{z}^{\zeta}\}}\right)(\eta)
		+
		 a_z\left(\mathcal{L}_z\mathbbm{1}_{\{\mathcal{I}_{z}^{\zeta}\}}\right)(\eta).\nonumber
	\end{eqnarray}
	Substituting \eqref{eq:16}, \eqref{eq:17}, \eqref{eq:18} in the above expression  we obtain
	\begin{eqnarray}
	\label{eq:2000}
		\left(\mathcal{L}\mathbbm{1}_{\{\mathcal{I}_{z}^{\zeta}\}}\right)(\eta) 
		&=&
		\sum_{y\;:\;(z,y)\in E}a_{z,y}\left(A^{\zeta}_{1}+\sum_{\delta=1}^{n}F_{+1}^{\zeta\delta}\mathbbm{1}_{\{\mathcal{I}_{y}^{\delta}\}}(\eta)+\sum_{\gamma=1}^{n}B_{1}^{\zeta\gamma}\mathbbm{1}_{\{\mathcal{I}_{z}^{\gamma}\}}(\eta)+\sum_{\gamma,\delta=1}^{n}G_{+1}^{\zeta\gamma\delta}\mathbbm{1}_{\{\mathcal{I}_{y}^{\gamma}\}}(\eta)\mathbbm{1}_{\{\mathcal{I}_{z}^{\delta}\}}(\eta)\right)\nonumber\\
		&+&
		\sum_{x\;:\;(x,z)\in E}a_{x,z}\left(A^{\zeta}_{2}+\sum_{\gamma=1}^{n}F_{-1}^{\zeta\gamma}\mathbbm{1}_{\{\mathcal{I}_{x}^{\gamma}\}}(\eta)+\sum_{\delta=1}^{n}C_{2}^{\zeta\delta}\mathbbm{1}_{\{\mathcal{I}_{z}^{\delta}\}}(\eta)+\sum_{\gamma,\delta=1}^{n}G_{-1}^{\zeta\gamma\delta}\mathbbm{1}_{\{\mathcal{I}_{z}^{\gamma}\}}(\eta)\mathbbm{1}_{\{\mathcal{I}_{x}^{\delta}\}}(\eta)\right)\nonumber\\
		&+&
		a_z\left(A^{\zeta}(z)+\sum_{\beta=1}^{n}F^{\zeta\beta}(z)\mathbbm{1}_{\{\mathcal{I}_{z}^{\beta}\}}(\eta)\right).
	\end{eqnarray}
	
	\subsection{Evolution equations}
	Using equation
	\eqref{eq:2000} for the right hand side of  \eqref{pippobaudo}
    we obtain the evolution equation for the average occupation.
    Recalling the notation
    in \eqref{ave} and \eqref{corr}
    (for the sake of space we do not write the explicit $t$-dependence) we arrive to
		\begin{eqnarray}
	\label{eq:20}
		\frac{d}{dt} \mu_z^{(\zeta)}
		&=&
		\sum_{y\;:\;(z,y)\in E}a_{z,y}\left(A^{\zeta}_{1}+\sum_{\delta=1}^{n}F_{+1}^{\zeta\delta}\;\mu_{y}^{(\delta)}+\sum_{\gamma=1}^{n}B_{1}^{\zeta\gamma}\;\mu_{z}^{(\gamma)}+\sum_{\gamma,\delta=1}^{n}G_{+1}^{\zeta\gamma\delta}\;c_{y,z}^{(\gamma,\delta)}\right)\nonumber\\
		&+&
		\sum_{x\;:\;(x,z)\in E}a_{x,z}\left(A^{\zeta}_{2}+\sum_{\gamma=1}^{n}F_{-1}^{\zeta\gamma}\;\mu_{x}^{(\gamma)}+\sum_{\delta=1}^{n}C_{2}^{\zeta\delta}\;\mu_{z}^{(\delta)}+\sum_{\gamma,\delta=1}^{n}G_{-1}^{\zeta\gamma\delta}\;c_{z,x}^{(\gamma,\delta)}\right)\nonumber\\
		&+&
		a_z\left(A^{\zeta}(z)+\sum_{\beta=1}^{n}F^{\zeta\beta}(z)\;\mu_{z}^{(\beta)}\right).
	\end{eqnarray}
	We notice that the equations for the time-dependent averages $\mu_z^{(\zeta)}(t)$ are not closed, as they involve the correlations $c_{z,z'}^{(\zeta,\zeta')}(t)$. 
	\begin{remark}[The process on the lattice]
	The generator \eqref{gen-gen} is an generalization of the lattice generator studied in \cite{wadati} to a general graph  with the addition of open boundaries. 
 Indeed, take as a special graph the $d$-dimensional regular lattice $\mathbb{Z}^{d}$ and ignore the boundaries. Then, calling
 $e^{(k)}$ the unit vector in the $k^{th}$ direction ($k=1,\ldots, d$) and defining
	\begin{equation}
		\begin{split}
		E^{\zeta}=A^{\zeta}_{1}+A^{\zeta}_{2}\\
		F_{0}^{\zeta\beta}=C_{2}^{\zeta\beta}+B_{1}^{\zeta\beta}
		\end{split}
	\end{equation}
equation \eqref{eq:2000} becomes	\begin{equation}
	\begin{split}
		\left(\mathcal{L}\mathbbm{1}_{\{\mathcal{I}_{z}^{\zeta}\}}\right)(\eta)=\sum_{k=1}^{d}\left\{E^{\zeta}+\sum_{\beta=1}^{n}\sum_{j=-1}^{+1}F_{j}^{\zeta\beta}\mathbbm{1}_{\{\mathcal{I}_{z+je^{(k)}}^{\beta}\}}(\eta)+\sum_{\beta,\beta^{'}=1}^{n}\sum_{j=\pm 1}G_{j}^{\zeta\beta\beta^{'}}\mathbbm{1}_{\{\mathcal{I}_{z+je^{(k)}}^{\beta}\}}(\eta)\mathbbm{1}_{\{\mathcal{I}_{z}^{\beta^{'}}\}}(\eta)\right\}
		\end{split}
	\end{equation}
	which is equation $(3.12)$ in \cite{wadati}.

	\end{remark}
	\section{Boundary-driven chains with linear reaction-diffusion}
	\label{sec4}
	In this and the following sections we specialize to the case with only two species, labelled by $1$ and $2$. Furthermore, we specialize to the one-dimensional geometry by considering a undirected linear chain.
	
	More precisely, the graph has $N$ vertices labelled by $\{1,2,\ldots,N\}$ with a distinguish role of the sites $\{1,N\}$which model two reservoirs. The interaction is of nearest neighbor type, i.e. 
	\begin{equation*}
	    a_{x,y}=\begin{cases}
	        1\qquad \text{if}\quad |x-y|=1\\
	        0\qquad \text{otherwise}
	    \end{cases}\qquad\qquad a_{x}=\begin{cases}
	        1\qquad \text{if}\quad x\in \{1,N\}\\
	        0\qquad \text{otherwise}
	    \end{cases}
	\end{equation*}
	%Each interior vertex $z=2,\ldots,N-1$ is adjacent to the previous one, i.e. $z-1$ and to the next one, i.e. $z+1$. The sites $1$ and $N$ are adjacent
	%just to sites $2$ and $N-1$, respectively, and in contact with two reservoirs: the ``left" reservoir is connected to site $1$ and the ``right" reservoir 
	%is connected to site $N$.
	It is convenient to call the sites $\{2,\ldots,N-1\}$ as \app bulk" and the two end sites $\{1,N\}$ as \app boundary". 
	The generator of the process thus reads as:
	\begin{equation}
		\mathcal{L}=\mathcal{L}_{1}+\sum_{z=1}^{N-1}\mathcal{L}_{z,z+1}+\mathcal{L}_{N}
	\end{equation}
	We specialize the result of Eq. \eqref{eq:20} to the boundary-driven chain. 
	Introducing $\forall \zeta,\beta=1,2$:
	\begin{align*}
		&F_{0}^{\zeta\beta}=B_{1}^{\zeta\beta}+C_{2}^{\zeta\beta}\quad
		&E^{\zeta}=A_{1}^{\zeta}+A_{2}^{\zeta}\\
		&A_{L}^{\zeta}=A^{\zeta}(1)\quad
		&A_{R}^{\zeta}=A^{\zeta}(N)\\
		&F^{\zeta\beta}_{L}=F^{\zeta\beta}(1)\quad
		&F^{\zeta\beta}_{R}=F^{\zeta\beta}(N)
	\end{align*}
	the evolution equations for the densities of the two species at site $z\in\{1,2,\ldots, N\}$ are given by:
	\begin{equation}\label{actionBDl}
		\begin{split}
			\frac{d}{dt} \mu_1^{(\zeta)}&=
			A_{L}^{\zeta}+A_{1}^{\zeta}+\sum_{\beta=1}^{2}\left(\left(B_{1}^{\zeta\beta}+F_{L}^{\zeta\beta}\right)\mu_{1}^{(\beta)}+F_{+1}^{\zeta\beta}\mu_{2}^{(\beta)}\right)\qquad\qquad\qquad\qquad\qquad\qquad\qquad\\
			&+\sum_{\beta,\beta^{'}=1}^{2}G_{+1}^{\zeta\beta\beta^{'}}c_{1,2}^{(\beta,\beta^{'})}
		\end{split}
	\end{equation}
	\noindent
	\begin{equation}\label{actionBU}
		\begin{split}
			\frac{d}{dt}\mu_z^{(\zeta)}&=E^{\zeta}+\sum_{\beta=1}^{2}\left(F_{-1}^{\zeta\beta}\mu_{z-1}^{(\beta)}+F_{0}^{\zeta\beta}\mu_{z}^{(\beta)}+F_{+1}^{\zeta\beta}\mu_{z+1}^{(\beta)}\right)\qquad\qquad\qquad \text{if } z\in \{2,\ldots,N-1\}\qquad\\&+ \sum_{\beta,\beta^{'}=1}^{2}\left(G_{-1}^{\zeta\beta\beta^{'}}c_{z-1,z}^{(\beta,\beta^{'})}+G_{+1}^{\zeta\beta\beta^{'}}c_{z,z+1}^{(\beta,\beta^{'})}\right\}
		\end{split}
	\end{equation}
	\begin{equation}\label{actionBDr}
		\begin{split}
			\frac{d}{dt}\mu_N^{(\zeta)}&=
			A_{R}^{\zeta}+A_{2}^{\zeta}+\sum_{\beta=1}^{2}\left(\left(C_{2}^{\zeta\beta}+F_{R}^{\zeta\beta}\right)\mu_{N}^{(\beta)}+F_{-1}^{\zeta\beta}\mu_{N-1}^{(\beta)}\right)\qquad\qquad\qquad\qquad\qquad\qquad\\
			&+\sum_{\beta,\beta^{'}=1}^{2}G_{-1}^{\zeta\beta\beta^{'}}c_{N-1,N}^{(\beta,\beta^{'})}
		\end{split}
	\end{equation}	
In the next section, we simplify the evolution equations for the average density by selecting a
subclass of processes with closed equations and
a linear structure.
\subsection{Imposing  the matching}
	\label{sec5}
	One could go further and compute the hierarchy of equations for higher-order correlation function \cite{wadati}.
	For general choices of the rate matrices $\Gamma$ and $W$, the equations do not close. In the following, we shall focus on those
	choices of rates that satisfy the following
	two requirements:
	\begin{enumerate}
	    \item \textit{Closure of the correlation equations}. This amounts to requiring that
	    the correlation terms in (\ref{actionBDl}), (\ref{actionBU}), (\ref{actionBDr})
	    vanish. It is shown in \cite{wadati}
	    that the vanishing of correlations
	    actually implies closure of the multi-point correlation function at all orders.
	    \item \textit{The average occupations follow the discretization of the reaction diffusion equation}. 
	    Considering the reaction diffusion system (\ref{stronglyWeaklycoupledPDE}), 
	    we approximate the laplacians with the central difference operators. We call
	    $\rho_i^{(\alpha)}$ the density of species $\alpha\in \{0,1,2\}$ at vertex $i\in \{1,\ldots, N\}$ with the constraint $\rho_{i}^{(0)}+\rho_{i}^{(1)}+\rho_{i}^{(2)}=1$.  Furthermore we fix the densities  at the left end (vertex 1) to the values of $\rho_L^{(1)}$, $\rho_L^{(2)}$ and similarly at the right end (vertex $N$) we impose $\rho_R^{(1)}$, $\rho_R^{(2)}$.   Then the discretization  of the two component reaction diffusion equations
(\ref{stronglyWeaklycoupledPDE}), reads as
\begin{equation}\label{goalBdl}
	\begin{split}
		\frac{d}{dt}\rho_{1}^{(1)}=&\sigma_{11}\left(\rho_{L}^{(1)}-2\rho^{(1)}_{1}+\rho^{(1)}_{2}\right)+\sigma_{12}\left(\rho_{L}^{(2)}-2\rho^{(2)}_{1}+\rho^{(2)}_{2}\right)+\Upsilon\left(\rho^{(2)}_{1}-\rho^{(1)}_{1}\right)\\
		\frac{d}{dt}\rho_{1}^{(2)}=&\sigma_{21}\left(\rho_{L}^{(1)}-2\rho^{(1)}_{1}+\rho^{(1)}_{2}\right)+\sigma_{22}\left(\rho_{L}^{(2)}-2\rho^{(2)}_{1}+\rho^{(2)}_{2}\right)+\Upsilon\left(\rho^{(1)}_{1}-\rho^{(2)}_{2}\right)
	\end{split}
\end{equation}

\begin{equation}\label{goalBulk}
	\begin{split}
		\frac{d}{dt}\rho_{z}^{(1)}&=\sigma_{11}\left(\rho^{(1)}_{z-1}-2\rho^{(1)}_{z}+\rho^{(1)}_{z+1}\right)+\sigma_{12}\left(\rho^{(2)}_{z-1}-2\rho^{(2)}_{z}+\rho^{(2)}_{z+1}\right)+\Upsilon\left(\rho^{(2)}_{z}-\rho^{(1)}_{z}\right)\\
		\frac{d}{dt}\rho_{z}^{(2)}&=\sigma_{21}\left(\rho^{(1)}_{z-1}-2\rho^{(1)}_{z}+\rho^{(1)}_{z+1}\right)+\sigma_{22}\left(\rho^{(2)}_{z-1}-2\rho^{(2)}_{z}+\rho^{(2)}_{z+1}\right)+\Upsilon\left(\rho^{(1)}_{z}-\rho^{(2)}_{z}\right)\\
		&\qquad\qquad\qquad\qquad\qquad\qquad\qquad\qquad\qquad\qquad\qquad\qquad\forall z=2,\ldots,N-1
	\end{split}
\end{equation}

\begin{equation}\label{goalBdr}
	\begin{split}
		\frac{d}{dt}\rho_{N}^{(1)}=&\sigma_{11}\left(\rho_{N-1}^{(1)}-2\rho^{(1)}_{N}+\rho^{(1)}_{R}\right)+\sigma_{12}\left(\rho_{N-1}^{(2)}-2\rho^{(2)}_{N}+\rho^{(2)}_{R}\right)+\Upsilon\left(\rho^{(2)}_{N}-\rho^{(1)}_{N}\right)\\
		\frac{d}{dt}\rho_{N}^{(2)}=&\sigma_{21}\left(\rho_{N-1}^{(1)}-2\rho^{(1)}_{N}+\rho^{(1)}_{R}\right)+\sigma_{22}\left(\rho_{N-1}^{(2)}-2\rho^{(2)}_{N}+\rho^{(2)}_{R}\right)+\Upsilon\left(\rho^{(1)}_{N}-\rho^{(2)}_{N}\right)
	\end{split}
\end{equation}
We  impose that the evolution equations for the averaged
occupations given in (\ref{actionBDl}), (\ref{actionBU}), (\ref{actionBDr})
do coincide with the discretized reaction-diffusion equations (\ref{goalBdl}), (\ref{goalBulk}), (\ref{goalBdr}).
	\end{enumerate}

\noindent
By imposing the closure condition 1.
and the discrete linear reaction-diffusion condition 2.
we get the set of equations described below.

\paragraph{Conditions from the bulk.} We first consider equation \eqref{actionBU}  which we require to have the form of (\ref{goalBulk}). We  obtain the following conditions:
\begin{itemize}
	\item \textit{Closure conditions}: equation (\ref{goalBulk}) has no second order terms, thus:
	\begin{equation}\label{closureConditions}
	G_{+1}^{\alpha\beta\beta^{'}}=0\qquad
			G_{-1}^{\alpha\beta\beta^{'}}=0\quad \forall \alpha,\beta,\beta^{'}=1,2
	\end{equation} 
	\begin{comment}
		This is accomplished by imposing the so called \app closure conditions":
		\begin{align}
			\sum_{\beta=0}^{2}\left(\Gamma^{\alpha\beta}_{\gamma\delta}+\Gamma_{00}^{\alpha\beta}\right)=&\sum_{\beta=0}^{2}\left(\Gamma_{\gamma 0}^{\alpha\beta}+\Gamma_{0\delta}^{\alpha\beta}\right)\;\;\,\;\;\;\;\forall \alpha,\gamma,\delta=1,2\;:\;\alpha\neq \gamma\label{CC1}\\
			\sum_{\beta=0}^{2}\left(\Gamma^{\beta\alpha}_{\gamma\delta}+\Gamma_{00}^{\beta\alpha}\right)=&\sum_{\beta=0}^{2}\left(\Gamma_{\gamma 0}^{\beta\alpha}+\Gamma_{0\delta}^{\beta\alpha}\right)\;\;\,\;\;\;\;\forall \alpha,\gamma,\delta=1,2\;:\;\alpha\neq \delta\label{CC2}\\
			\sum_{\beta=0}^{2}\left(\sum_{j=0\;:\;j\neq \alpha}\Gamma_{\alpha 0}^{j\beta}+\Gamma_{00}^{\alpha\beta}\right)=&\sum_{\beta=0}^{2}\left(\sum_{j=0\;:\;j\neq \alpha}\Gamma_{\alpha \delta}^{j\beta}+\Gamma_{0\delta}^{\alpha\beta}\right)\;\;\,\;\;\;\;\forall \alpha,\delta=1,2\label{CC3}\\
			\sum_{\beta=0}^{2}\left(\sum_{j=0\;:\;j\neq \alpha}\Gamma_{ 0\alpha}^{\beta j}+\Gamma_{00}^{\beta\alpha}\right)=&\sum_{\beta=0}^{2}\left(\sum_{j=0\;:\;j\neq \alpha}\Gamma_{\gamma \alpha}^{j\beta}+\Gamma_{\gamma 0}^{\alpha\beta}\right)\;\;\,\;\;\;\;\forall \alpha,\gamma=1,2\label{CC4}
		\end{align}
		Each of the equations from (\ref{CC1}) to (\ref{CC4}) gives $4$ equations that must be fulfilled. In total, these requirements give $16$ conditions. 
	\end{comment}
	The above requirement leads to $16$ conditions on the transition rates $\Gamma_{\gamma\delta}^{\alpha\beta}$.
	\item \textit{Laplacian conditions}: the one point correlation function should evolve as the coupled discrete Laplacian in (\ref{goalBulk}) with linear reaction. This is accomplished by imposing:
	\begin{align}
\label{Laplacianconditions}
			&F_{-1}^{11}=F_{+1}^{11}=\sigma_{11}&
			&F_{-1}^{12}=F_{+1}^{12}=\sigma_{12}&
			&F_{-1}^{21}=F_{+1}^{21}=\sigma_{21}&\nonumber
			&F_{-1}^{22}=F_{+1}^{22}=\sigma_{22}\\
			&F_{0}^{11}=-2\sigma_{11}-\Upsilon&
			&F_{0}^{12}=-2\sigma_{12}+\Upsilon&
			&F_{0}^{21}=-2\sigma_{21}+\Upsilon&
			&F_{0}^{22}=-2\sigma_{22}-\Upsilon&
\end{align}

	\begin{comment}
	\begin{equation}\label{Laplacianconditions}
		\begin{cases}
			F_{-1}^{11}=F_{+1}^{11}=\sigma_{11}\\
			F_{-1}^{12}=F_{+1}^{12}=\sigma_{12}\\
			F_{-1}^{21}=F_{+1}^{21}=\sigma_{21}\\
			F_{-1}^{22}=F_{+1}^{22}=\sigma_{22}\\
			F_{0}^{11}=-2\sigma_{11}-\Upsilon\\
			F_{0}^{12}=-2\sigma_{12}+\Upsilon\\
			F_{0}^{21}=-2\sigma_{21}+\Upsilon\\
			F_{0}^{22}=-2\sigma_{22}-\Upsilon
		\end{cases}
	\end{equation}
	\end{comment}
	The above requirement leads to $12$ conditions on the transition rates $\Gamma_{\gamma\delta}^{\alpha\beta}$.
	\item \textit{Zero-order terms}: equation (\ref{goalBulk}) has no zero-order term, thus: 
	\begin{equation}\label{annulmentKnownTerm}
E^{1}=0\qquad E^{2}=0
	\end{equation}
	The above requirement leads to $2$ conditions on the transition rates $\Gamma_{\gamma\delta}^{\alpha\beta}$.\\
\end{itemize}
Our task is to determine the $81$ transition rates $\Gamma_{\gamma\delta}^{\alpha\beta}$ $\forall \alpha,\beta,\gamma,\delta=0,1,2$ that define  the bulk infinitesimal generator.
By exploiting the stochasticity properties of the generator
(sum of the elements on the rows must be zero), the problem reduces to finding $72$ transition rates. By considering (\ref{closureConditions}), (\ref{Laplacianconditions}), (\ref{annulmentKnownTerm}), only $16+12+2=30$ conditions are available. This means that the problem to solve is under-determined.

For the analysis that will follow, it is convenient to introduce an unknown vector $\mathbf{u}\in \mathbb{R}_{+}^{72}$ that contains the desired $72$ transition rates, and an appropriate matrix ${K}\in \mathbb{R}^{30\times 72}$ and  vector $\mathbf{b}\in \mathbb{R}^{30}$. Then, it is possible (for details see Appendix \ref{appendice}) to rewrite  (\ref{closureConditions}), (\ref{Laplacianconditions}), (\ref{annulmentKnownTerm}) as: 
\begin{equation}\label{LinearBIGsystem}
	K\mathbf{u}=\mathbf{b}.
\end{equation}
The matrix $K$ is full rank, thus there exists a family of solutions with $42$ free parameters. Furthermore we have to guarantee the non-negativity of the solution, as the transition rates
are non-negative.
For later use, recalling the definitions of $F,G,E$'s,
we observe that  the conditions (\ref{closureConditions}), (\ref{Laplacianconditions}), (\ref{annulmentKnownTerm}) actually only involve sums of three transition rates.

\paragraph{Conditions from the boundaries.} We now want to find conditions to match (\ref{actionBDl}) and  (\ref{actionBDr}) with (\ref{goalBdl}) and (\ref{goalBdr}), respectively. We consider the conditions on the left boundary; the right boundary is treated similarly. We get:
\begin{itemize}
	\item \textit{Closure conditions}: the vanishing of correlation in (\ref{actionBDl}) is already guaranteed by (\ref{closureConditions}).
	\item \textit{Laplacian conditions}: 
		\begin{align*}
			&F_{L}^{11}+B_{1}^{11}=-2\sigma_{11}-\Upsilon&
			&F_{L}^{12}+B_{1}^{12}=-2\sigma_{12}+\Upsilon&
			&F_{+1}^{11}=\sigma_{11}&
			&F_{+1}^{12}=\sigma_{12}\\
			&F_{L}^{22}+B_{1}^{22}=-2\sigma_{22}-\Upsilon&
			&F_{L}^{21}+B_{1}^{21}=-2\sigma_{21}+\Upsilon&
			&F_{+1}^{21}=\sigma_{21}&
			&F_{+1}^{22}=\sigma_{22}&
	\end{align*}
	Since the equations that involve $F_{+1}^{\zeta,\delta}$ are already imposed in (\ref{Laplacianconditions}), inserting the definition of the $F_{L}^{\zeta,\delta}$, the above conditions reduce to 
	\begin{align}\label{LaplacianiBoundary}
		&-W_{0}^{1}(1)-W_{1}^{0}(1)-W_{1}^{2}(1)+B_{1}^{11}=-2\sigma_{11}-\Upsilon&\nonumber
		&B_{1}^{12}+W_{2}^{1}(1)-W_{0}^{1}(1)=-2\sigma_{12}+\Upsilon\\
		&W_{1}^{2}(1)-W_{0}^{2}(1)+B_{1}^{21}=-2\sigma_{21}+\Upsilon&
		&-W_{2}^{0}(1)-W_{0}^{2}(1)-W_{2}^{1}(1)+B_{1}^{22}=-2\sigma_{22}-\Upsilon
\end{align}
	\item \textit{Zero-order terms}: 
	\begin{equation*}
	A_{L}^{1}+A^{1}_{1}=\sigma_{11}\rho_{L}^{(1)}+\sigma_{12}\rho_{L}^{(2)}\qquad
			A_{L}^{2}+A^{2}_{1}=\sigma_{21}\rho_{L}^{(1)}+\sigma_{22}\rho_{L}^{(2)} 
	\end{equation*}
	As a consequence of (\ref{annulmentKnownTerm}), $A_{2}^{\zeta}$ are zero. Therefore, the above conditions reduce to  
	\begin{equation}\label{termineNotoBoudary}	W_{0}^{1}(1)=\sigma_{11}\rho_{L}^{(1)}+\sigma_{12}\rho_{L}^{(2)}\qquad
			W_{0}^{2}(1)=\sigma_{21}\rho_{L}^{(1)}+\sigma_{22}\rho_{L}^{(2)} 
	\end{equation}  
\end{itemize}
All in all, combining (\ref{LaplacianiBoundary}) and (\ref{termineNotoBoudary}) we see that the rates of the
boundary generators are uniquely determined
 by the bulk rates. Indeed, for a choice of the bulk rates (which in turn appear in the $B_1^{\zeta,\delta}$), we have:
%\begin{equation}\label{BoundarySystemL}
%\begin{cases}
%	W_{0}^{1}(1)=\sigma_{11}\rho_{L}^{(1)}+\sigma_{12}\rho_{L}^{(2)}\\
%	W_{0}^{2}(1)=\sigma_{21}\rho_{L}^{(1)}+\sigma_{22}\rho_{L}^{(2)}\\
%	-W_{0}^{1}(1)-W_{1}^{0}(1)-W_{1}^{2}(1)+B_{1}^{11}=-2\sigma_{11}-\Upsilon\\
%	B_{1}^{12}+W_{2}^{1}(1)-W_{0}^{1}(1)=-2\sigma_{12}+\Upsilon\\
%	W_{1}^{2}(1)-W_{0}^{2}(1)+B_{1}^{21}=-2\sigma_{21}+\Upsilon\\
%	-W_{2}^{0}(1)-W_{0}^{2}(1)-W_{2}^{1}(1)+B_{1}^{22}=-2\sigma_{22}-\Upsilon
%	\end{cases}
%	\end{equation}
\begin{align}\label{BoundarySystemL}
		&W_{0}^{1}(1)=\sigma_{11}\rho_{L}^{(1)}+\sigma_{12}\rho_{L}^{(2)}&\nonumber
		&W_{0}^{2}(1)=\sigma_{21}\rho_{L}^{(1)}+\sigma_{22}\rho_{L}^{(2)}\\ 
		&W_{0}^{1}(1)+W_{1}^{0}(1)+W_{1}^{2}(1)=2\sigma_{11}+\Upsilon+B_{1}^{11}&
		&W_{2}^{1}(1)-W_{0}^{1}(1)=-2\sigma_{12}+\Upsilon-B_{1}^{12}\\ \nonumber
		&W_{1}^{2}(1)-W_{0}^{2}(1)=-2\sigma_{21}+\Upsilon-B_{1}^{21}&
		&W_{2}^{0}(1)+W_{0}^{2}(1)+W_{2}^{1}(1)=2\sigma_{22}+\Upsilon+B_{1}^{22}
\end{align}
On the right boundary, a similar argument yields: 
\begin{align}\label{BoundarySystemR}
		&W_{0}^{1}(N)=\sigma_{11}\rho_{R}^{(1)}+\sigma_{12}\rho_{R}^{(2)}&\nonumber
		&W_{0}^{2}(N)=\sigma_{21}\rho_{R}^{(1)}+\sigma_{22}\rho_{R}^{(2)}\\
		&W_{0}^{1}(N)+W_{1}^{0}(N)+W_{1}^{2}(N)=2\sigma_{11}+\Upsilon+C_{2}^{11}&
		&W_{2}^{1}(N)-W_{0}^{1}(N)=-2\sigma_{12}+\Upsilon-C_{2}^{12}\\
		&W_{1}^{2}(N)-W_{0}^{2}(N)=-2\sigma_{21}+\Upsilon-C_{2}^{21}&
		&W_{2}^{0}(N)+W_{0}^{2}(N)+W_{2}^{1}(N)=2\sigma_{22}+\Upsilon+C_{2}^{22}&\nonumber
\end{align}
Let us notice that (\ref{BoundarySystemL}) and (\ref{BoundarySystemR}) are determined systems of algebraic equations in the unknowns $W^{\cdot}_{\cdot}(1),W^{\cdot}_{\cdot}(N)$.

\subsection{Determination of the rates}
\label{sec6}
\label{two-par}
Our first main result is contained in Theorem \ref{THM_principal}.
It identifies a necessary and sufficient condition (in terms of two parameters $h,m\ge 0$) on the diffusivity matrix $\Sigma$ and the reaction coefficient $\Upsilon$ such that the one-di\-men\-sio\-nal boundary
driven chain with two-species has averaged densities satisfying the discrete linear reaction-diffusion
equations (\ref{goalBdl}), (\ref{goalBulk}), (\ref{goalBdr}). 
Furthermore, by setting $h=m$, it provides the example of a one-parameter family of {\em symmetric} models with such a property.
To state the example it is convenient to 
introduce the  \textit{mutation map} $\alpha \mapsto \bar{\alpha}$ defined by:
\begin{equation}\label{mutationMap}
\begin{split}
	&1\rightarrow 2\\
	&2\rightarrow 1\\
	&0\rightarrow 0\,.
\end{split}
\end{equation} 

\begin{theorem}\label{THM_principal}
Let $\Sigma$ be a $2\times 2$ positive definite diffusion
matrix and $\Upsilon>0$ be a reaction coefficient. Let $\rho_{L}^{(1)}$ and $\rho_{L}^{(2)}$
(respectively, $\rho_{R}^{(1)}$ and $\rho_{R}^{(2)}$) be the densities of the species $1$ and $2$
at the left (respectively, right) boundary.
Then, for any choice of $h,m\geq 0$ there exist boundary-driven interacting particle systems on the chain $\{1,\ldots,N\}$ such that their evolution equations of the average occupation variable are (\ref{goalBdl}), (\ref{goalBulk}), (\ref{goalBdr}) if and only if the diffusion matrix coefficients $\sigma_{11},\sigma_{12},\sigma_{21},\sigma_{22}$ and the reaction coefficient $\Upsilon$ are non-negative and fulfill the conditions
\begin{equation}\label{conditionTHM}
	\sigma_{11}+\sigma_{21}=\sigma_{12}+\sigma_{22}\qquad
	\sigma_{12}\leq \frac{\Upsilon-m}{2}\qquad
	\sigma_{21}\leq \frac{\Upsilon-h}{2}\,.
\end{equation}
%Consider a fixed reaction diffusion model of type (\ref{stronglyWeaklycoupledPDE}) (defined by diffusivity matrix with coefficients $\sigma_{11},\sigma_{12},\sigma_{21}$, $\sigma_{22}\in \mathbb{R}$, the reaction constant $\Upsilon \in \mathbb{R}$ and the density boundary values $\rho_{L}^{(1)},\rho_{L}^{(2)},$ $\rho_{R}^{(1)},\rho_{R}^{(2)}\in \mathbb{R}^{+}$). Then, there exists a two parameter family of interacting particle systems, spanned by $h,m\geq 0$, such that their evolution equations of the average occupation variable are (\ref{goalBdl}), (\ref{goalBulk}), (\ref{goalBdr}) under the following constraints
%\begin{equation}\label{conditionTHM}
%\begin{cases}
%	\sigma_{11},\sigma_{12},\sigma_{21},\sigma_{22},\Upsilon\geq 0\\
%	\sigma_{11}+\sigma_{21}=\sigma_{12}+\sigma_{22}\\
%	\sigma_{12}\leq \frac{\Upsilon-m}{2}\\
%	\sigma_{21}\leq \frac{\Upsilon-h}{2}
%\end{cases}
%\end{equation}
%and 
%\begin{equation}\label{conditionsBDC}
%\begin{cases}
%	0\leq \rho_{L}^{(1)}+\rho_{L}^{(2)}\leq 1\\
%	0\leq \rho_{R}^{(1)}+\rho_{R}^{(2)}\leq 1
%\end{cases}
%\end{equation}
Moreover, an explicit example of a symmetric generator (parameterized by $h= m\ge 0$) is given by
	\begin{equation}
		{L}={L}_{1}+\sum_{x=1}^{N-1}{L}_{x,x+1}+{L}_{N}
	\end{equation}
	with edge generator
\begin{eqnarray}
\label{example1par}
{L}_{x,x+1} f(\eta) 
&= & 
\sigma_{11} (f(\eta_1,\ldots,\eta_{x+1},\eta_{x},\ldots,\eta_N)-f(\eta)) \nonumber \\
&+&
\sigma_{12} (f(\eta_1,\ldots,\bar{\eta}_{x+1},\bar{\eta}_{x},\ldots,\eta_N)-f(\eta)) \nonumber \\
&+&
(\Upsilon - 2\sigma_{12} -m)(f(\eta_1,\ldots,\bar{\eta}_{x},\eta_{x+1},\ldots,\eta_N)-f(\eta))\nonumber \\
&+&
m(f(\eta_1,\ldots,\eta_{x},\bar{\eta}_{x+1},\ldots,\eta_N)-f(\eta))\,.
\end{eqnarray}
The site generator at the left boundary
is given by
\begin{eqnarray}
\label{genboundary1par}
	{L}_{1} f(\eta) 
	&= & (\sigma_{11}\rho_{L}^{(1)}+\sigma_{12}\rho_{L}^{(2)})\mathbbm{1}_{\{\mathcal{I}_{1}^{0}\}}(\eta)\left[f(\eta_{1}+\delta^{1},\ldots,\eta_{N})-f(\eta_{1},\ldots,\eta_{N})\right]\nonumber\\
	&+&(\sigma_{12}\rho_{L}^{(1)}+\sigma_{11}\rho_{L}^{(2)})\mathbbm{1}_{\{\mathcal{I}_{1}^{0}\}}(\eta)\left[f(\eta_{1}+\delta^{2},\ldots,\eta_{N})-f(\eta_{1},\ldots,\eta_{N})\right]\nonumber\\
	&+&(\sigma_{11} + \sigma_{12})\rho_{L}^{(0)} \mathbbm{1}_{\{\mathcal{I}_{1}^{1}\}}(\eta)\left[f(\eta_{1}-\delta^{1},\ldots,\eta_{N})-f(\eta_{1},\ldots,\eta_{N})\right]\nonumber \\
		&+&
		(\sigma_{11} + \sigma_{12})\rho_{L}^{(0)} \mathbbm{1}_{\{\mathcal{I}_{1}^{2}\}}(\eta)\left[f(\eta_{1}-\delta^{2},\ldots,\eta_{N})-f(\eta_{1},\ldots,\eta_{N})\right]\nonumber\\
	&+&(m + \sigma_{12}\rho_{L}^{(1)} + \sigma_{11}\rho_{L}^{(2)})\mathbbm{1}_{\{\mathcal{I}_{1}^{1}\}}(\eta)\left[f(\eta_{1}+\delta^{2}-\delta^{1},\ldots,\eta_{N})-f(\eta_{1},\ldots,\eta_{N})\right]\nonumber \\
	&+&
	(m+\sigma_{11}\rho_{L}^{(1)} + \sigma_{12}\rho_{L}^{(2)})\mathbbm{1}_{\{\mathcal{I}_{1}^{2}\}}(\eta)\left[f(\eta_{1}-\delta^{2}+\delta^{1},\ldots,\eta_{N})-f(\eta_{1},\ldots,\eta_{N})\right]
\end{eqnarray}
where $\rho_{L}^{(0)}:=1-\rho_{L}^{(1)}-\rho_{L}^{(2)}$ . Here $\pm\delta^{\alpha}$ denotes the addition/removal of species $\alpha$. 
The site generator at the right boundary
is defined similarly (now with parameters
$\rho_R^{(1)}$ and $\rho_R^{(2)}$).
\end{theorem}
Before discussing the proof of the theorem,
a few comments are collected in the following
remarks.
\begin{remark}
\label{corollarioDisaccoppiato}\textnormal{
The theorem is in agreement with the previous literature results stating that in the absence of the reaction term, for the existence of the two dimensional coupled heat equations the cross diffusivities must vanish (\cite{gorban2011quasichemical}, \cite{kahane}). Here we find the corresponding statement at the level of the particle process. Indeed, by assuming $\Upsilon=0$, then 
the condition \eqref{conditionTHM} can be 
satisfied iff
%\begin{equation*}
%\begin{cases}
%	\sigma_{12}=\sigma_{21}=h=m=0\\
%	\sigma_{11}=\sigma_{22}\geq 0
%\end{cases}
%\end{equation*}
%\end{corollary}
%Indeed, by applying Theorem \ref{THM_principal} and the assumption $\Upsilon=0$:
%\begin{equation*}
%\begin{cases}
%	\sigma_{11},\sigma_{12},\sigma_{21},\sigma_{22},h,m\geq 0\\
%	\sigma_{11}+\sigma_{21}=\sigma_{12}+\sigma_{22}\\
%	\sigma_{12}\leq \frac{-m}{2}\\
%	\sigma_{21}\leq\frac{-h}{2}
%\end{cases}
%\end{equation*}
%Thus, the only way to fulfill all the above equations is 
$\sigma_{12}=\sigma_{21}=h=m=0$ and $\sigma_{11}=\sigma_{22}$. }
\end{remark}
\begin{remark}\textnormal{
The transitions allowed by the edge generator (\ref{example1par}) are the following: 
	\begin{equation}\label{allowedTransitions}
		(\gamma,\delta)\quad\rightarrow\quad\begin{cases}
			(\delta,\gamma) \qquad\text{stirring at rate $\sigma_{11}$}\\
			(\overline{\delta},\overline{\gamma})\qquad\text{stirring and mutation at rate $\sigma_{12}$}\\
			(\overline{\gamma},\delta)\qquad\text{left mutation at rate $\Upsilon - 2\sigma_{12} -m$}\\
			(\gamma,\overline{\delta})\qquad\text{right mutation at rate $m$}\\
		\end{cases}
		%\qquad \forall \gamma,\delta=0,1,2
	\end{equation}
Thus we see that the rate of stirring is associated to the diffusion coefficient $\sigma_{11}$, while
the rate of stirring with mutation is related  to the cross-diffusion coefficient $\sigma_{12}$. The rates of the left and right
mutations are precisely tuned to guarantee
that, for all $m\ge0$, the evolution equations of the average occupation variables are (\ref{goalBdl}), (\ref{goalBulk}), (\ref{goalBdr}). A visual representation of this process is showed in Figure \ref{fig:LINE}. In particular,
the choice $m=0$ kills the right mutations,
the choice $m=\Upsilon - 2\sigma_{12}$
kills the left mutations, while the choice
$m=\frac{\Upsilon}{2} - \sigma_{12}$
gives the same rate to left and right mutations.
Let us also observe that only when $m=0$, the boundary generators satisfy the conditions $\forall z\in \{1,N\}$: 
\begin{equation}
        W_{1}^{0}(z)=W_{2}^{0}(z)\qquad
        W_{0}^{1}(z)=W_{2}^{1}(z)\qquad
        W_{0}^{2}(z)=W_{1}^{2}(z)\,.
    \end{equation}}
    \begin{figure}
        \centering
        \includegraphics[scale=0.6]{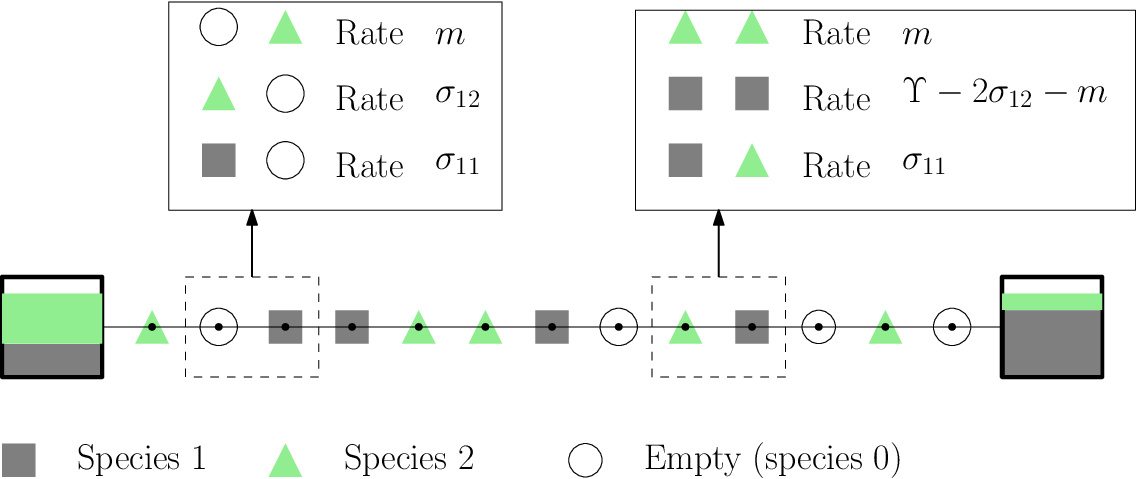}
        \caption{The boundary driven process with generator  \eqref{example1par}, \eqref{genboundary1par} .
    Grey squares identify species 1, green triangles species 2, and white circles the empty state. The reservoirs are represented by rectangles, where the interior colours denote the particles or vacuum densities. In the boxes, we give two examples of allowed bulk transition with the corresponding rates.}
        \label{fig:LINE}
    \end{figure}
%    as required in \cite{vanicat}. 
%
%\textbf{Let us also notice that the transitions defined in (\ref{allowedTransitions}) never create or annihilate particles. In other words, the total mass of a certain bond is unchanged by transition. This seems to tell that a macroscopic continuity equation of the sum of the two species should hold.}\\
\end{remark}
\begin{remark}\label{remarkFamily} \textnormal{
It is possible to exhibit a particle process with a generator
having the same structure of \eqref{example1par}
but containing two parameters $h,m\ge 0$
and depending on all the coefficients of the diffusivity matrix $\sigma_{11},\sigma_{12},\sigma_{21},\sigma_{22}$
and on the reaction coefficient $\Upsilon$, provided they fulfill
condition \eqref{conditionTHM}. This is shown in
Appendix \ref{appendixa}. When $h\neq m$
the matrix associated to the generator 
$L_{x,x+1}$ is generically not symmetric and the four transitions described in \eqref{allowedTransitions}
have rates which depend on the
specific configuration values.
When $h=m$ the generator
$L_{x,x+1}$ is symmetric if the diffusivity matrix is, i.e  
$\sigma_{12}=\sigma_{21}$,
and thus as a consequence of \eqref{conditionTHM} 
the elements on the diagonal are equal, i.e. $\sigma_{11}=\sigma_{22}$.}
\end{remark}
\begin{remark}\label{Remark_colorblind}
\textnormal{Considering the ``color-blind'' process, i.e. the process that does not distinguish between the particles of type 1 and those of type 2, we obtain a process with just occupied or empty sites. This is indeed the
classical boundary-driven simple symmetric exclusion process \cite{derrida1993exact}, where in the bulk particles jump to the left or to the right at rate $\sigma:=\sigma_{11}+\sigma_{12}$, provided
there is space,
and at the left boundary particles are
created at rate $\sigma\rho_L$ and removed
at rate $\sigma(1-\rho_L)$, where $\rho_L$ is
the particle density (and similarly at the right boundary with density $\rho_R$).} 
\end{remark}
%With such a choice, the generator is symmetric, but still depends on two non physical parameters $h,m$. One might want to remove the dependence on $h,m$, in order to obtain one generator for a fixed reaction diffusion system.  One could fix $h=m=0$. In this case, these two parameters do not appear in the generator, however the "right mutation" transition will disappear.
%To keep all the possible transitions and remove the dependence on $h,m$ it could be convenient to chose $h,m$ as follows
%\begin{equation}
%    \begin{cases}
%        h=\frac{\Upsilon}{2}-\sigma_{21}\\
%        m=\frac{\Upsilon}{2}-\sigma_{12}
%    \end{cases}
%\end{equation}
%Here, the generator is still symmetric when the physical model is (i.e. when the physical constants satisfy (\ref{simmetriaModelloFisico})), and  the right and the left mutations occur with the same rate. \\

\medskip
\noindent
{\bf Proof of Theorem \ref{THM_principal}.}
We provide here the main ideas; full details of the proof are given in the appendix \ref{appendice}. We first consider the bulk part and then the boundary one.
\begin{itemize}
\item {\em Bulk process}: To find the rates of the bulk process we need to solve (\ref{LinearBIGsystem}), i.e. the system $K\mathbf{u}=\mathbf{b}$ where $K$ is a matrix of size $30\times 72$
and $\mathbf{b}$ is a vector described in the appendix \ref{appendice}. 
This system has a great under-determination order (72-30=42).
To overcome this difficulty, we exploit the fact that, as already noticed in the text following \eqref{LinearBIGsystem}, the required conditions (\ref{closureConditions}), (\ref{Laplacianconditions}), (\ref{annulmentKnownTerm}) only involve sums of three rates. 
As a consequence,
we may introduce a new system where
the unknowns are the summed triples.
This new system, which will be denoted by
$\Xi \mathbf{y}=\mathbf{b}$ 
where $\Xi$ is a matrix of size $30\times 36$, has an under-determination order equal to 6, and thus can be solved explicitly under the non-negativity constraint on $\mathbf{y}$ (see Appendix \ref{appendixa}). It is precisely the request $\mathbf{y}\ge 0$ that 
further reduces the under-determination
order to $2$ (parametrized by the parameters $h,m \ge 0$) and produces the constraint \eqref{conditionTHM}.

Once the vector $\mathbf{y}$, whose components are sum of three rates, has been found, the next step is the identification of the transition rates themselves. This of course can be
done in several ways. To produce an
explicit example we have followed
the two criteria below:
\begin{itemize}
	%\item \textit{Non negativity of the rates}: i.e $\mathbf{u} \geq 0$
	\item {The matrix associated to the generator has the greatest number of zeros.}
	\item {Choice of the following rates}: \begin{equation}
			\Gamma_{12}^{21}=\sigma_{11}\qquad
			\Gamma_{21}^{12}=\sigma_{22}\qquad
			\Gamma_{11}^{22}=\sigma_{21}\qquad
			\Gamma_{22}^{11}=\sigma_{12}.
	\end{equation}
\end{itemize}
After simple but long computations, this choice leads to the generator
(\ref{explicitGeneratoBulk}) in Appendix \ref{appendixa} involving the two parameters $h,m\ge 0$. When
we set $h=m$ and we choose a symmetric diffusivity matrix (which in turn guarantees a symmetric particle process) the generator \eqref{example1par} is obtained.
%
%{\color{blue} This process has many vanishing rates and preserves a certain "homogeneity" in the transitions that keeps a more physical meaning, and allows to include the stirring process in this family of Markovian models.  Furthermore, when the parameters $h,m$ vanish, the solution with the above constraints coincides with the least square one.}
\item{\em Boundary process}: to find the rates of the boundary process we need to solve (\ref{BoundarySystemL}) and (\ref{BoundarySystemR}). Having already determined the rates of the bulk process, by direct computation we find the boundary generators (\ref{leftGeneral}) and (\ref{rightGeneral}) reported in the appendix \ref{appendixa}, which depend on $h,m\ge 0$. When
we set $h=m$ and choose a symmetric diffusivity matrix, then the generator \eqref{genboundary1par} is obtained.
\end{itemize}
\begin{flushright} $\Box$ \end{flushright}

\section{Duality and hydrodynamic limit}\label{dualitySection}
We aim to derive the hydrodynamic equations for the family of processes defined in \eqref{example1par}. In this section, we assume to work on the whole one-dimensional lattice $\mathbb{Z}$. To formulate the results, it is convenient to change notation. 
	The state space of the Markov process defined by the edge generator \eqref{example1par} on the full line can be identified with the three-dimensional simplex $$\widetilde{\Omega}=\left\{(n_{0},n_{1},n_{2})\in \{0,1\}^{3}\;:\; n_{0}+n_{1}+n_{2}=1\right\}^{\mathbb{Z}}\,.$$ 
	In this notation, the component $n^{z}$ at site $z\in\mathbb{Z}$ of a configuration $n\in\widetilde{\Omega}$ is thus a triplet with two 0's and a 1, whose position is associated with a hole, or with a particle of type 1, or with a particle of type 2. For example, $(n_{0}^{z},n_{1}^{z},n_{2}^{z})=(0,1,0)$ indicates that in the site $z\in \mathbb{Z}$ there is one particle of species 1. 
	Then, recalling the notation in \eqref{mutationMap} for the mutation map, the process $\{n(t) , t\ge 0\}$ taking values in $\widetilde{\Omega}$ is defined
	by the following generator $L$ working of local functions $f:\widetilde{\Omega}\to\mathbb{R}$:
	\begin{align}\label{bulkDuableGeneratore}
	    &{L}=\sum_{z\in\mathbb{Z}}L_{z,z+1}
	    \end{align}
with
	    \begin{align*}
	    L_{z,z+1}&= \sigma_{11}L_{z,z+1}^{S}+\sigma_{12}L_{z,z+1}^{SM}+(\Upsilon-2\sigma_{12}-m)L_{z,z+1}^{LM}+m L_{z,z+1}^{RM}
	    \end{align*}
	    where
	    \begin{align}\label{linearOperators}
	    L_{z,z+1}^{S}f(n)=&\sum_{\alpha,\beta=0}^{2}n^{z}_{\alpha}n^{z+1}_{\beta}\left[f(n-\delta_{\alpha}^{z}+\delta_{\beta}^{z}+\delta_{\alpha}^{z+1}-\delta_{\beta}^{z+1})-f(n)\right]\nonumber
	    \\
	    L_{z,z+1}^{SM}f(n)=&
	    \sum_{\alpha,\beta=0}^{2}n_{\alpha}^{z}n_{\beta}^{z+1}\left[f(n-\delta_{\alpha}^{z}+\delta_{\overline{\beta}}^{z}-\delta_{\beta}^{z+1}+\delta_{\overline{\alpha}}^{z+1})-f(n)\right]\nonumber
	    \\
	    L_{z,z+1}^{LM}f(n)=&
	    \sum_{\alpha=0}^{2}n_{\alpha}^{z}\left[f(n-\delta_{\alpha}^{z}+\delta_{\overline{\alpha}}^{z})-f(n)\right]\nonumber
	    \\
	    L_{z,z+1}^{RM}f(n)=&
	    \sum_{\beta=0}^{2}n_{\beta}^{z+1}\left[f(n-\delta_{\beta}^{z+1}+\delta_{\overline{\beta}}^{z+1})-f(n)\right]
	\end{align}
	A fundamental tool for the hydrodynamic limit is duality:
 usually, the hydrodynamic limit is dictated by the scaling properties of one dual particles. We say that the Markov process with generator \eqref{bulkDuableGeneratore}
	is self-dual with respect to the 
	self-duality function $D:\widetilde{\Omega} \times\widetilde{\Omega} \to \mathbb{R}$
	if for all $t\ge 0$ and for all $(n,\ell)\in \widetilde{\Omega}\times \widetilde{\Omega}$
	$$
	\mathbb{E}_n[D(n(t),\ell)] = \mathbb{E}_\ell[D(n,\ell(t))]
	$$
	where on the left hand side $\mathbb{E}_n$ denotes expectation
	in the process $\{n(t), t \ge 0\}$ initialized from the configuration $n$
	and, analogously, on the right hand side $\mathbb{E}_\ell$ denotes expectation
	in $\{\ell(t), t \ge 0\}$ which is a copy of the process initialized from the configuration $\ell$.
	
	In this section, by abuse of notation, we denote 	$\mathbbm{1}_{\{a\geq b \}}$
	the function defined by 
	\begin{equation*}
	   \mathbbm{1}_{\{a\geq b \}}= \begin{cases}
	        1\qquad \text{ if}\quad a\geq b\\
	        0\qquad \text{ if}\quad a< b
	    \end{cases}
	\end{equation*}
	\begin{theorem}[Self-Duality]
	The Markov process $\{n(t), t \ge 0\}$ defined by the generator \eqref{bulkDuableGeneratore} is self-dual with the self duality function
		\begin{equation}\label{dualityfunction}
	D(n,\ell)=\prod_{z\in\mathbb{Z}}\prod_{k=1}^{2}\mathbbm{1}_{\{n_{k}^{z}\geq \ell_{k}^{z} \}}
\end{equation}
	\end{theorem} 
{\bf Proof:}
	It is enough to prove that 
	\begin{equation}\label{Duality}
		\left({L}D(\cdot,\ell)\right)(n)=\left({L}D(n,\cdot)\right)(\ell)\qquad \forall (n, \ell) \in \widetilde{\Omega}\times \widetilde{\Omega}
	\end{equation}
The generator \eqref{bulkDuableGeneratore} is a superposition of four generators. Remarkably, the duality relation can be verified for each of them. Indeed, one has:
\begin{align*}
&({L}_{z,z+1}^{S}D(\cdot,\ell))(n)
\\&=
\left[\mathbbm{1}_{\{n_{1}^{z+1}\geq \ell_{1}^{z}\}}\mathbbm{1}_{\{n_{2}^{z+1}\geq \ell_{2}^{z}\}}\mathbbm{1}_{\{n_{1}^{z}\geq \ell_{1}^{z+1}\}}\mathbbm{1}_{\{n_{2}^{z}\geq \ell_{2}^{z+1}\}}-\mathbbm{1}_{\{n_{1}^{z}\geq
\ell_{1}^{z}\}}\mathbbm{1}_{\{n_{2}^{z}\geq \ell_{2}^{z}\}}\mathbbm{1}_{\{n_{1}^{z+1}\geq \ell_{1}^{z+1}\}}\mathbbm{1}_{\{n_{2}^{z+1}\geq \ell_{2}^{z+1}\}}\right] \prod_{x\notin{\{z,z+1\}}}\prod_{k=1}^{2}\mathbbm{1}_{\{n_{k}^{x}\geq \ell_{k}^{x} \}} 
\\ &=
\left[\mathbbm{1}_{\{n_{1}^{z}\geq \ell_{1}^{z+1}\}}\mathbbm{1}_{\{n_{2}^{z}\geq \ell_{2}^{z+1}\}}\mathbbm{1}_{\{n_{1}^{z+1}\geq \ell_{1}^{z}\}}\mathbbm{1}_{\{n_{2}^{z+1}\geq \ell_{2}^{z}\}}-\mathbbm{1}_{\{n_{1}^{z}\geq \ell_{1}^{z}\}}\mathbbm{1}_{\{n_{2}^{z}\geq \ell_{2}^{z}\}}\mathbbm{1}_{\{n_{1}^{z+1}\geq \ell_{1}^{z+1}\}}\mathbbm{1}_{\{n_{2}^{z+1}\geq \ell_{2}^{z+1}\}}\right]\prod_{x\notin{\{z,z+1\}}}\prod_{k=1}^{2}\mathbbm{1}_{\{n_{k}^{x}\geq \ell_{k}^{x} \}}\\ &
=(L_{z,z+1}^{S}D(n,\cdot))(\ell)\,.
\end{align*}
Similarly, one has
\begin{align*}
&({L}_{z,z+1}^{SM}D(\cdot,\ell))(n)
\\&=
\left[\mathbbm{1}_{\{n_{2}^{z+1}\geq \ell_{1}^{z}\}}\mathbbm{1}_{\{n_{1}^{z+1}\geq \ell_{2}^{z}\}}\mathbbm{1}_{\{n_{2}^{z}\geq \ell_{1}^{z+1}\}}\mathbbm{1}_{\{n_{1}^{z}\geq \ell_{2}^{z+1}\}}-\mathbbm{1}_{\{n_{1}^{z}\geq \ell_{1}^{z}\}}\mathbbm{1}_{\{n_{2}^{z}\geq \ell_{2}^{z}\}}\mathbbm{1}_{\{n_{1}^{z+1}\geq \ell_{1}^{z+1}\}}\mathbbm{1}_{\{n_{2}^{z+1}\geq\ \ell_{2}^{z+1}\}}\right]\prod_{x\notin{\{z,z+1\}}}\prod_{k=1}^{2}\mathbbm{1}_{\{n_{k}^{x}\geq \ell_{k}^{x} \}}
\\&=
\left[\mathbbm{1}_{\{n_{1}^{z}\geq \ell_{2}^{z+1}\}}\mathbbm{1}_{\{n_{2}^{z}\geq \ell_{1}^{z+1}\}}\mathbbm{1}_{\{n_{1}^{z+1}\geq \ell_{2}^{z}\}}\mathbbm{1}_{\{n_{2}^{z+1}\geq \ell_{1}^{z}\}}-\mathbbm{1}_{\{n_{1}^{z}\geq \ell_{1}^{z}\}}\mathbbm{1}_{\{n_{2}^{z}\geq \ell_{2}^{z}\}}\mathbbm{1}_{\{n_{1}^{z+1}\geq \ell_{1}^{z+1}\}}\mathbbm{1}_{\{n_{2}^{z+1}\geq \ell_{2}^{z+1}\}}\right]\prod_{x\notin{\{z,z+1\}}}\prod_{k=1}^{2}\mathbbm{1}_{\{n_{k}^{x}\geq \ell_{k}^{x} \}}
\\&=
L_{z,z+1}^{SM}(D(n,\cdot)(\ell)\,.
\end{align*}
For the generator that mutates at site $z$ we have
\begin{align*}
	({L}_{z,z+1}^{LM}D(\cdot,\ell))(n)&=
	\left[\mathbbm{1}_{\{n_{2}^{z}\geq \ell_{1}^{z}\}}\mathbbm{1}_{\{n_{1}^{z}\geq \ell_{2}^{z}\}}-\mathbbm{1}_{\{n_{1}^{z}\geq \ell_{1}^{z}\}}\mathbbm{1}_{\{n_{2}^{z}\geq \ell_{2}^{z}\}}\right]\prod_{x\neq z}\prod_{k=1}^{2}\mathbbm{1}_{\{n_{k}^{x}\geq \ell_{k}^{x} \}}\\&=
	\left[\mathbbm{1}_{\{n_{1}^{z}\geq \ell_{2}^{z}\}}\mathbbm{1}_{\{n_{2}^{z}\geq \ell_{1}^{z}\}}-\mathbbm{1}_{\{n_{1}^{z}\geq \ell_{1}^{z}\}}\mathbbm{1}_{\{n_{2}^{z}\geq \ell_{2}^{z}\}}\right]\prod_{x\neq z}\prod_{k=1}^{2}\mathbbm{1}_{\{n_{k}^{x}\geq \ell_{k}^{x} \}}\\&=
	(L_{z,z+1}^{LM}D(n,\cdot))(\ell)\,,
\end{align*}
and analogously, for the generator that mutates at site $z+1$, we find
\begin{align*}
	({L}_{z,z+1}^{RM}D(\cdot,\ell))(n)
	&=
	\left[\mathbbm{1}_{\{n_{2}^{z+1}\geq \ell_{1}^{z+1}\}}\mathbbm{1}_{\{n_{1}^{z+1}\geq \ell_{2}^{z+1}\}}-\mathbbm{1}_{\{n_{1}^{z+1}\geq \ell_{1}^{z+1}\}}\mathbbm{1}_{\{n_{2}^{z+1} \geq \ell_{2}^{z+1}\}}\right]\prod_{x\neq z+1}\prod_{k=1}^{2}\mathbbm{1}_{\{n_{k}^{x}\geq \ell_{k}^{x} \}}
	\\&=
	\left[\mathbbm{1}_{\{n_{1}^{z+1}\geq \ell_{2}^{z+1}\}}\mathbbm{1}_{\{n_{2}^{z+1}\geq \ell_{1}^{z+1}\}}-\mathbbm{1}_{\{n_{1}^{z+1}\geq \ell_{1}^{z+1}\}}\mathbbm{1}_{\{n_{2}^{z+1}\geq \ell_{2}^{z+1}\}}\right]\prod_{x\neq z+1}\prod_{k=1}^{2}\mathbbm{1}_{\{n_{k}^{x}\geq \ell_{k}^{x}\}}\\ &=
	(L_{z,z+1}^{RM}D(n,\cdot)(\ell)
\end{align*}
\begin{flushright}
$\Box$\end{flushright}

\medskip

To formulate the hydrodynamic limit, we consider
a scaling parameter $\epsilon \ge 0$ and we  introduce the empirical density fields
%	\begin{equation}
%	    X_{1}^{N}(t)=\frac{1}{N}\sum_{z\in \mathbb{Z}}\mathbbm{1}_{\{\mathcal{I}_{z}^{1}(t N^2)\}}\delta_{z/N}\quad  X_{2}^{N}(t)=\frac{1}{N}\sum_{z\in \mathbb{Z}}\mathbbm{1}_{\{\mathcal{I}_{x}^{2}(t N^2)\}}\delta_{z/N}
%	\end{equation}
	\begin{equation}\label{densityField}
	    X_{1}^{\epsilon}(t)=\epsilon\sum_{z\in \mathbb{Z}}n_{1}^{z}( \epsilon^{-2}t)\delta_{\epsilon z}\,\qquad X_{2}^{\epsilon}(t)=\epsilon\sum_{z\in \mathbb{Z}}n_{
	    2}^{z}(\epsilon^{-2}t)\delta_{\epsilon z}
	\end{equation}
The empirical density fields $\{X_1^{\epsilon}(t), t\ge 0\}$ and
$\{X_2^{\epsilon}(t), t\ge 0\}$ are measure-valued processes constructed from the process $\{n(t), t\ge 0\}$. We also need to specify a good set of initial distributions.
\begin{definition}
Let $\widehat{\rho}^{(\alpha)}: \mathbb{R}\to [0,1]$, with $\alpha\in \{1,2\}$, be a continuous bounded real function called the initial macroscopic profile. A sequence $(\mu_{\epsilon})_{\epsilon \ge 0}$ of measures on $\widetilde{\Omega}$, is a sequence of compatible initial conditions if
$\forall\alpha\in \{1,2\}$, $\forall \delta>0$: 
    \begin{equation}
        \lim_{\epsilon\rightarrow 0}\mu_{\epsilon}\left(\left|\langle X_{\alpha}^{\epsilon}(0),g\rangle-\int_{\mathbb{R}}g(x)\widehat{\rho}^{(\alpha)}(x)dx\right|>\delta\right)=0
    \end{equation}
    where $g:\mathbb{R}\to\mathbb{R}$ is a smooth test function with compact support.
	\end{definition}
	We then have the following theorem for the hydrodynamic limit.
\begin{theorem}[Hydrodynamic limit of the Markov process $\{n(t), t \ge 0\}$].\label{HDTheorem_fixed}
    Let $\widehat{\rho}^{(\alpha)}$ with $\alpha\in\{1,2\}$ be initial macroscopic profiles and $(\mu_{\epsilon})_{\epsilon> 0}$ be a sequence of compatible initial conditions. Let $\mathbb{P}_{\mu_{\epsilon}}$ be the law of the measure valued process $(X_{1}^{\epsilon}(t),X_{2}^{\epsilon}(t))$ defined in \eqref{densityField}. Then $\forall T,\delta>0$,$\forall \alpha\in \{1,2\}$ and for all smooth test function with compact support $g:\mathbb{R}\to\mathbb{R}$
    \begin{equation}\label{HDstatement_fixed}
    \lim_{\epsilon\to 0}\mathbb{P}_{\mu_{\epsilon}}\left(\sup_{t\in [0,T]}\left|\langle X_{\alpha}^{\epsilon}(t),g\rangle-\int_{\mathbb{R}}g(x)\rho^{(\alpha)}(x,t)dx\right|>\delta\right)=0,
\end{equation}
where $\rho^{(1)},\rho^{(2)}$ are the strong solutions of 
    \begin{equation}\label{HDLimit_fix}
        \begin{cases}
\partial_{t}\rho^{(1)}=\sigma_{11}\partial_{x}^2\rho^{(1)}+\widetilde{\Upsilon}\left(\rho^{(2)}-\rho^{(1)}\right)\\
	\partial_{t}\rho^{(2)}=\sigma_{11}\partial_{x}^2\rho^{(2)}+\widetilde{\Upsilon}\left(\rho^{(1)}-\rho^{(2)}\right)\\	\rho^{(\alpha)}(0,x)=\widehat{\rho}^{(\alpha)}(x)\qquad \forall x\in [0,1],\;\forall \alpha\in\{1,2\}            
        \end{cases}
    \end{equation}
\end{theorem}
\textbf{Proof}: The proof is standard and it is based on the Dynkin's martingale and its quadratic variation. For the tightness and the uniqueness of the limiting point we refer to \cite{presutti} and \cite{seppalainen2008translation}. we provide here some details for the computations of the Dynkin's martingale and its quadratic variation via Carré-Du-Champ. 

 We introduce the following real and positive parameters: 
 \begin{equation}
 \label{scaling}
 \widetilde{\sigma}_{12}=\epsilon^{-2}\sigma_{12}, \quad \widetilde{\Upsilon}=\epsilon^{-2}\Upsilon \quad 
 \widetilde{m}=\epsilon^{-2}m.
 \end{equation} 
 We consider the re-scaled generator
 \begin{equation}
     L^{(\epsilon)}=\sum_{z\in \mathbb{Z}}L_{z,z+1}^{(\epsilon)}
 \end{equation} where 
\begin{equation}\label{rescaledGenerator}
    L^{(\epsilon)}_{z,z+1}= \sigma_{11}L_{z,z+1}^{S}+\widetilde{\sigma}_{12}\epsilon^{2}L_{z,z+1}^{SM}+\epsilon^{2}(\widetilde{\Upsilon}-2\widetilde{\sigma}_{12}-\widetilde{m})L_{z,z+1}^{LM}+\widetilde{m}\epsilon^{2} L_{z,z+1}^{RM}.
\end{equation}
By choosing  $\forall z\in \mathbb{Z}$ and $\forall \alpha\in\{1,2\}$ the action of the rescaled generator on $n_{\alpha}^{z}$ is the following: 
\begin{align*}
    (L^{(\epsilon)}n_{\alpha}^{x})(n)=\sigma_{11}\left(n_{\alpha}^{z+1}-2n_{\alpha}^{z}+n_{\alpha}^{z-1}\right)+\widetilde{\sigma}_{12}\epsilon^{2}\left(n_{\overline{\alpha}}^{z+1}-2n_{\alpha}^{z}+n_{\overline{\alpha}}^{z-1}\right)+\epsilon^{2}\left(\widetilde{\Upsilon}-2\widetilde{\sigma}_{12}\right)\left(n_{\overline{\alpha}}^{z}-n_{\alpha}^{z}\right)
\end{align*}
By consequence considering a test function $g$
\begin{align*}
    \int_{0}^{t }ds\; \epsilon^{-2} L^{(\epsilon)}\langle X_{\alpha}^{\epsilon}(s),g\rangle&=\sigma_{11}\int_{0}^{t}ds\; \epsilon^{-2}\;\epsilon\sum_{z\in\mathbb{Z}}n_{\alpha}^{z}(s)\left[g\left((z+1)\epsilon\right)-2g\left(z\epsilon\right)+g\left((z-1)\epsilon\right)\right]
    \\&+
    \widetilde{\sigma}_{12}\int_{0}^{t}ds\;\epsilon^{-2}\epsilon^{3}\; \sum_{z\in\mathbb{Z}}\left(n_{\overline{\alpha}}^{z}(s)\left[g\left((z+1)\epsilon\right)+g\left((z-1)\epsilon\right)\right]-2n_{\alpha}^{z}(s)g(z\epsilon) \right)
    \\&+
    \int_{0}^{t}ds\;\epsilon^{-2}\;\epsilon^{3}(\widetilde{\Upsilon}-2\widetilde{\sigma}_{12})\sum_{z\in\mathbb{Z}}g\left(z\epsilon\right)\left[n_{\overline{\alpha}}^{z}-n_{\alpha}^{z}\right]
\end{align*}
By using the Taylor expansion we rewrite the above equality as
\begin{align*}
    \int_{0}^{t }ds\; \epsilon^{-2}L^{(\epsilon)}\langle X_{\alpha}^{\epsilon}(s),g\rangle&=\sigma_{11}\int_{0}^{t}\epsilon \sum_{z\in\mathbb{Z}}n_{\alpha}^{z}\Delta g\left(z\epsilon\right)+\widetilde{\sigma}_{12}\int_{0}^{t}\epsilon^{3} \sum_{z\in\mathbb{Z}}n_{\overline{\alpha}}^{z}\Delta g\left(z\epsilon\right)+\widetilde{\Upsilon}\int_{0}^{t}\epsilon\sum_{z\in\mathbb{Z}}g\left(z\epsilon\right)\left[n_{\overline{\alpha}}^{z}-n_{\alpha}^{z}\right]+o(\epsilon)\\&=\sigma_{11}\int_{0}^{t}\epsilon \sum_{z\in\mathbb{Z}}n_{\alpha}^{z}\Delta g\left(z\epsilon\right)+\widetilde{\Upsilon}\int_{0}^{t}\epsilon\sum_{z\in\mathbb{Z}}g\left(z\epsilon\right)\left[n_{\overline{\alpha}}^{z}-n_{\alpha}^{z}\right]+o(\epsilon).
\end{align*}
Defining the Dynkin's martingale $\forall \alpha\in\{1,2\}$ 
\begin{equation}
    M_{g}^{t}(X_{\alpha}^{\epsilon}):=\langle X_{\alpha}^{\epsilon}(t),g\rangle-\langle X_{\alpha}^{\epsilon}(0),g\rangle-\int_{0}^{t}\epsilon^{-2}L^{(\epsilon)}\langle X_{\alpha}^{\epsilon}(s),g\rangle ds,
\end{equation}
by the previous computations, we have
\begin{align*}
     M_{g}^{t}(X^{\epsilon}_{\alpha})+o(\epsilon)=\langle X_{\alpha}^{\epsilon}(t),g\rangle-\langle X_{\alpha}^{\epsilon}(0),g\rangle-\sigma_{11}\int_{0}^{t}\langle X_{\alpha}^{\epsilon}(s),\Delta g\rangle ds-\widetilde{\Upsilon}\int_{0}^{t}\langle X_{\overline{\alpha}}^{\epsilon}(s)-X_{\alpha}^{\epsilon}(s),g\rangle ds.
\end{align*}
The right-hand side is the discrete counterpart of the weak solution of \eqref{HDLimit_fix}. \\
To have tightness of the law of the measure-valued processes \eqref{densityField} we need to show that 
\begin{equation}
    \lim_{\epsilon\to 0}\mathbb{E}_{\mu_{\epsilon}}\left[M_{g}^{t}(X_{\alpha}^{\epsilon})^{2}\right]=0.
\end{equation}
We first observe that 
\begin{align*}
    \mathbb{E}_{\mu_{\epsilon}}\left[M_{g}^{t}(X_{\alpha}^{\epsilon})^{2}\right]\leq \mathbb{E}_{\mu_{\epsilon}}\left[\sup_{t\in [0,T]}|M_{g}^{t}(X_{\alpha}^{\epsilon})|^{2}\right]\leq 4\mathbb{E}_{\mu_{\epsilon}}\left[M_{g}^{T}(X_{\alpha}^{\epsilon})^{2}\right]=4\mathbb{E}_{\mu_{\epsilon}}\left[\int_{0}^{T}\epsilon^{-2}\Gamma_{g}^{s}(X_{\alpha}^{\epsilon})ds\right],
\end{align*}
where $\Gamma_{g}^{s}(X_{\alpha}^{\epsilon})$ is the Carré-Du-Champ operator that can be written as
\begin{equation}
\begin{split}
    \Gamma_{g}^{s}(X_{\alpha}^{\epsilon})&=L^{(\epsilon)}\langle  X_{\alpha}(t),g\rangle^{2}-2\langle X_{\alpha}(t),g\rangle L^{(\epsilon)}\langle X_{\alpha}(t),g\rangle.%+L_{\epsilon}^{SM}\langle  X_{\alpha}(t),g\rangle^{2}-2\langle X_{\alpha}(t),g\rangle L^{SM}_{\epsilon}\langle X_{\alpha}(t),g\rangle\\&+
   % L_{\epsilon}^{L}\langle  X_{\alpha}(t),g\rangle^{2}-2\langle X_{\alpha}(t),g\rangle L^{LM}_{\epsilon}\langle X_{\alpha}(t),g\rangle+
    %L_{\epsilon}^{RM}\langle  X_{\alpha}(t),g\rangle^{2}-2\langle X_{\alpha}(t),g\rangle L^{RM}_{\epsilon}\langle X_{\alpha}(t),g\rangle
    \end{split}
    \end{equation}
    By using the definition of the re-scaled generator \eqref{rescaledGenerator} we obtain the following
    \begin{align*}
        \epsilon^{-2}\Gamma_{g}^{s}(X_{\alpha}^{\epsilon})&=\sigma_{11}\epsilon^{2}\sum_{z\in\mathbb{Z}}\left[n_{\alpha}^{z}(1-n_{\alpha}^{z+1})+n_{\overline{\alpha}}^{z}(1-n_{\overline{\alpha}}^{z+1})\right]\left(\nabla g(z\epsilon)\right)^{2}\\&+\widetilde{\sigma}_{12}\epsilon^{2}\sum_{z\in\mathbb{Z}}\left\{2\left[n_{\alpha}^{z}n_{\overline{\alpha}}^{z+1}+n_{\overline{\alpha}}^{z}n_{\alpha}^{z+1}\right]g(z\epsilon)g((z+1)\epsilon)+n_{\overline{\alpha}}^{z}\left[g((z+1)\epsilon)^{2}+g((z-1)\epsilon)^{2}\right]+n_{\alpha}^{z}2g(z\epsilon)^{2}\right\}\\&+\left(\widetilde{\Upsilon}-2\widetilde{\sigma}_{12}\right)\epsilon^{2}\sum_{z\in\mathbb{Z}}(n_{\overline{\alpha}}^{z}+n_{\alpha}^{z})g(z\epsilon)^{2}+o(\epsilon^{2}).
    \end{align*}
    Let's introduce the set $\mathcal{S}_{g}$ as the smallest compact subset of $\mathbb{R}$ that contains the supports of a fixed $g$ and of the first two derivatives. Then, $|\mathcal{S}_{g}|\leq C^{'}\epsilon^{-1}$, with a $C^{'}$ positive and finite constant. Moreover, by the hard-core constraint $n_{\alpha}^{z}\leq 1$, $\forall z\in \mathbb{Z}$ and $\forall \alpha\in \{1,2\}$. By consequence, exploiting the smoothness of $g$ we derive the following bound
    \begin{equation}
        \mathbb{E}_{\mu_{\epsilon}}\left[\int_{0}^{T}\epsilon^{-2}\Gamma_{g}^{s}(X_{\alpha}^{\epsilon})ds\right]\leq C\epsilon,
    \end{equation}
    with $C<\infty$. This concludes the proof. 
   % By taking the limit we have
    %\begin{equation}
     %   \lim_{\epsilon\to 0}\mathbb{E}_{\mu_{\epsilon}}\left[(M_{\alpha,\phi}^{t})^{2}\right]\leq \lim_{\epsilon \to 0}C\epsilon=0
    %\end{equation}
    \begin{flushright}
        $\square$
    \end{flushright}
    %{\color{blue} Can we rescale the $\sigma_{12}$ in the $L^{SM}$ by $\epsilon^{3}$??}
\begin{remark}
    Let's define a ``color-blind" density field \begin{equation}
        X^{\epsilon}(t):=\epsilon\sum_{z\in\mathbb{Z}}n^{z}(t\epsilon^{-2})\delta_{z\epsilon}
    \end{equation}
    where ${n}^{z}(t):=n_{\alpha}^{z}(t)+n_{\overline{\alpha}}^{z}(t)$. By re-scaling only the $L^{RM}_{z,z+1}$ and $L_{z,z+1}^{LM}$ terms of the generator, the same proof of Theorem \ref{HDTheorem_fixed} we would give, as limiting PDE, the heat equation
    \begin{equation}
        \begin{cases}
            \partial_{t}\rho(x,t)=(\sigma_{11}+\sigma_{12})\partial_{xx}\rho(x,t)\\
            \rho(x,0)=\rho_{0}(x)
        \end{cases}
    \end{equation}
    This is in agreement with the Remark \ref{Remark_colorblind}. %and with the fact that we do not see global uphill. 
\end{remark}
\begin{remark}\label{remarknaive}
We observe that in order to obtain the hydrodynamic limit of the process
$\{n(t) ; t\ge 0\}$ we had to scale the parameters as in \eqref{scaling}. The `naive' scaling where the diffusivity parameter $\sigma_{11}$ and $\sigma_{12}$ are both kept constant (while the reaction parameters are scaled as $\Upsilon =  \epsilon^2 \tilde\Upsilon$ and $m = \epsilon^2 \tilde m$) is not viable, as it would lead to a violation of the maximum principle. Indeed, if we assume that the limiting PDEs 
are of the form
\begin{equation}
    \begin{cases}
        \partial_{t}\rho^{(\alpha)}=A\rho^{(\alpha)}\qquad \forall x\in [0,1],\;\forall \alpha\in\{1,2\}\\	\rho^{(\alpha)}(0,x)=\widehat{\rho}^{(\alpha)}(x)
    \end{cases}
\end{equation}
where the operator $A$ is defined as \begin{equation}
A\rho^{(\alpha)}:=\sigma_{11}\partial_{xx}\rho^{(\alpha)}+\sigma_{12}\partial_{xx}\rho^{(\overline{\alpha)}}+\widetilde{\Upsilon}\left(\rho^{(\overline{\alpha})}-\rho^{(\alpha)}\right)
\end{equation}
then $A$ does not satisfy the maximum principle. Indeed, it is possible to construct smooth functions $f^{(\alpha)}:\mathbb{R}\to \mathbb{R}$ such that, calling
\begin{equation}\label{max}
    f^{(\alpha)}(x_{*}^{(\alpha)}):=\max_{x\in \mathbb{R}} f^{(\alpha)}(x)
\end{equation} 
one obtains 
\begin{equation}\label{violationMP}
    Af^{(\alpha)}(x_{*}^{(\alpha)})=\sigma_{11}\partial_{xx} f^{(\alpha)}(x_{*}^{(\alpha)})+\sigma_{12}\partial_{xx}f^{(\overline{\alpha})}(x_{*}^{(\alpha)})+\widetilde{\Upsilon}\left(f^{(\overline{\alpha)}}(x_{*}^{(\alpha)})-f^{(\alpha)}(x_{*}^{(\alpha)})\right)>0.
\end{equation}
This follows by observing that \eqref{max} guarantees that $\partial_{xx}f^{(\alpha)}(x^{*})\leq 0$, but the other terms of the right hand side of \eqref{violationMP} can be positive and arbitrary large. As a consequence of the
violation of the maximum principle it follows that
$A$ can not be the generator of a Markov process.
From the microscopic point of view,
the problem with the `naive' rescaling is that
the rate of left mutations
\begin{equation}
    (\widetilde{\Upsilon}\epsilon^{2}-2\sigma_{12}-\widetilde{m}\epsilon^{2})
\end{equation}
becomes negative (!) for sufficiently small $\epsilon$.
\end{remark}
\begin{remark}
    If we perform the hydrodynamic limit with an ``Euler" re-scaling, i.e. we re-scale the time only by a factor $\epsilon$ and we define $\widehat{\sigma}_{12}=\epsilon^{-1}\sigma_{12}$, $\widehat{\Upsilon}=\epsilon^{-1}\Upsilon$ and $\widetilde{m}=\epsilon^{-1}m$ we obtain the following ODE's system
    \begin{equation}
        \begin{cases}
            \frac{d}{dt}\rho^{(1)}(t)=\widehat{\Upsilon}(\rho^{(2)}-\rho^{(1)})\\
            \frac{d}{dt}\rho^{(2)}(t)=\widehat{\Upsilon}(\rho^{(1)}-\rho^{(2)})\\
            \rho^{(1)}(0)=\rho^{(1)}_{0},\quad \rho^{(2)}(0)=\rho^{(2)}_{0}
        \end{cases}
    \end{equation}
    that is a purely reacting system. The ODE's are linear and the solution is given by
    \begin{equation}
        \begin{cases}
            \rho^{(1)}(t)=\frac{\rho^{(1)}_{0}+\rho^{(2)}_{0}}{2}+\frac{\rho^{(1)}_{0}-\rho^{(2)}_{0}}{2}e^{-2\widehat{\Upsilon}t}\\
            \rho^{(2)}(t)=\frac{\rho^{(1)}_{0}+\rho^{(2)}_{0}}{2}-\frac{\rho^{(1)}_{0}-\rho^{(2)}_{0}}{2}e^{-2\widehat{\Upsilon}t}
        \end{cases}
    \end{equation}
\end{remark}

\section{Conclusions}\label{sec8}
We considered  multi-species stochastic interacting particle systems with hard-core interaction defined on an directed graph. We also added site-generators, that allow to define the boundary-driven version having non-zero  stationary currents.

For a one dimensional chain with two species, we established that in order to have that the
average occupation evolves as the discrete counterpart of the linear reaction-diffusion equation \eqref{stronglyWeaklycoupledPDE}, the diffusivity matrix $\Sigma$ and the reaction coefficient
$\Upsilon$ have to fulfill  condition \eqref{conditionTHM} of Theorem \ref{THM_principal}.
As an additional  result, we have identified a one-parameter family of multi-species interacting particle systems (the one defined by the generator \eqref{example1par}) where the analysis can
be pushed further. In particular, due to the existence of a dual process, the hydrodynamic limit is deduced.
In the hydrodynamic regime the coupling between species
due to the cross-diffusivity coefficients disappears. 
The origin of this is that if
the cross-diffusivities are not scaled to zero
then the Markov property is lost (see Remark \ref{remarknaive}).
Partial uphill diffusion, although present
in a finite size system, is lost in the
hydrodynamic limit.

It would be interesting to extend
the analysis to a higher number of species.
As observed in \cite{krishna} the uphill phenomenology
of systems with three species of particles or more can be substantially different from the ones with two species.
Another open problem is the study of uphill diffusion
for systems with a {\em non-linear} reaction-diffusion structure,
i.e. with diffusivity matrix whose
elements are functions of the particle densities
\cite{quastel}. 
Finally, we mention that the family of models with
generator \eqref{example1par} includes the stirring process which is known to posses the algebraic
structure of the ${GL}(n)$ group (which in fact leads
to integrability of the model \cite{casiniRouvenGiardina}).
It would be interesting to check if the model we have introduced preserves such algebraic structure.
%
%with open boundaries, that has been proved to be integrable %(in the sense of quantum inverse scattering method) \footnote{See \cite{vanicat}}. By consequence, it could be natural to ask under which conditions the family of models defined in this paper keeps on satisfying this integrability property.\\
%Moreover, it has been proved in \cite{carinci2013duality}, %that the exclusion process (also with hard core interaction) with open boundaries, has an absorbing dual. Since, the stirring process seems to be the natural extension to the many type case of the SSEP, one could wonder under which conditions the family of models of Theorem \ref{THM_principal} posses an absorbing dual (or self dual) process. \\
%Furthermore, we restricted our analysis to the evolution equations of the average occupation variable. It could be interesting to rigorously proof the hydrodynamic limit and the fluctuations field of the described models. \\
%Finally, one might think to remove the hard-core exclusion rule and to generalize these kind of processes to higher spin models. 

\section*{Acknowledgment}
We thank Frank Redig for useful discussions and comments on the hydrodynamic limit.
This project has been funded under the National Recovery and Resilience Plan (NRRP),
Mission 04 Component 2 Investment 1.5 – NextGenerationEU, Call for tender
n. 3277 dated 30/12/2021.
Award Number:  0001052 dated 23/06/2022.
Research supported in part by GNFM-INdAM.
\newline 

\section*{Appendix}

\begin{appendices}
%\textbf{\Large{Appendices}}
\section{Steady state partial uphill diffusion}
\label{sec7}

Let us consider the steady state of (\ref{stronglyWeaklycoupledPDE}), with Dirichlet boundary conditions: 
\begin{equation}\label{stadystateEQN}
	\begin{split}
		&\sigma_{11}\frac{d^{2}}{dx^{2}} \rho^{(1)}(x)+\sigma_{12}\frac{d^{2}}{dx^{2}} \rho^{(2)}(x)+\Upsilon(\rho^{(2)}(x)-\rho^{(1)}(x))=0\\
		&\sigma_{21}\frac{d^{2}}{dx^{2}} \rho^{(1)}(x)+\sigma_{22}\frac{d^{2}}{dx^{2}} \rho^{(2)}(x)+\Upsilon(\rho^{(1)}(x)-\rho^{(2)}(x))=0 \\
		&\rho^{(1)}(0)=\rho_{L}^{(1)}\;\;\;\rho^{(2)}(0)=\rho_{L}^{(2)}\quad
		\rho^{(1)}(1)=\rho_{R}^{(1)}\;\;\;\rho^{(2)}(1)=\rho_{R}^{(2)}
	\end{split}
\end{equation}	

Recalling that diffusivity matrix (\ref{DiffusionMatrix}) is assumed to be positive definite we introduce the constants $A=\Upsilon\frac{\sigma_{12}+\sigma_{22}}{\sigma_{11}\sigma_{22}-\sigma_{12}\sigma_{21}}> 0$ and $B=-\Upsilon\frac{\sigma_{11}+\sigma_{21}}{\sigma_{11}\sigma_{22}-\sigma_{12}\sigma_{21}}< 0$. 
The solution of the above system of ordinary differential equations is 	
%\begin{equation}\label{solutionODE}
	%\begin{split}
	%\rho^{(1)}(x)&=E+Fx+C\left(1-2A\right)e^{-%\sqrt{2A}x}+D\left(1-2A\right)e^{\sqrt{2A}x}\\
	%	\rho^{(2)}(x)&=E+Fx+Ce^{-\sqrt{2A}x}+De^{\sqrt{2A}x}
%	\end{split}
%\end{equation}
\begin{equation}\label{solutionODE}
	\begin{split}
	\rho^{(1)}(x)&=E+Fx+C\left(1+\frac{A-B}{B}\right)e^{-\sqrt{A-B}x}+D\left(1+\frac{A-B}{B}\right)e^{\sqrt{A-B}x}\\
		\rho^{(2)}(x)&=E+Fx+Ce^{-\sqrt{A-B}x}+De^{\sqrt{A-B}x}	\end{split}
\end{equation}
where the constants $C,D,E,F $ are determined by the boundary conditions as follows: 
%\begin{align*}
		%&E=\frac{\,\rho_{L}^{(2)}+\,\rho_{L}^{(1)}}{2}&
	%	&C=	-\frac{\left(\rho_{L}^{(1)}\,e^{2\,\sqrt{2A}}-\rho_{L}^{(2)}\,e^{2\,\sqrt{2A}}-\rho_{R}^{(1)}\,e^{\sqrt{2A}}+\rho_{R}^{(2)}\,e^{\sqrt{2A}}\right)}{2\,\left(e^{2\,\sqrt{2A}}-1\right)}\\
%	&F=	-\frac{\,\rho_{L}^{(2)}-\,\rho_{R}^{(2)}+\,\rho_{L}^{(1)}-\,\rho_{R}^{(1)}}{2}&
	%	&D=-\frac{\left(\rho_{L}^{(1)}-\rho_{L}^{(2)}-\rho_{R}^{(1)}\,e^{\sqrt{2A}}+\rho_{R}^{(2)}\,e^{\sqrt{2A}}\right)}{2-2\,e^{2\,\sqrt{2A}}}
%\end{align*}
 
\begin{align*}
		&E=\frac{A\,\rho_{L}^{(2)}-B\,\rho_{L}^{(1)}}{A-B}&
		&C=	\frac{B\,\left(\rho_{L}^{(1)}\,e^{2\,\sqrt{A-B}}-\rho_{L}^{(2)}\,e^{2\,\sqrt{A-B}}-\rho_{R}^{(1)}\,e^{\sqrt{A-B}}+\rho_{R}^{(2)}\,e^{\sqrt{A-B}}\right)}{\left(A-B\right)\,\left(e^{2\,\sqrt{A-B}}-1\right)}\\
	&F=	-\frac{A\,\rho_{L}^{(2)}-A\,\rho_{R}^{(2)}-B\,\rho_{L}^{(1)}+B\,\rho_{R}^{(1)}}{A-B}&
		&D=\frac{B\,\left(\rho_{L}^{(1)}-\rho_{L}^{(2)}-\rho_{R}^{(1)}\,e^{\sqrt{A-B}}+\rho_{R}^{(2)}\,e^{\sqrt{A-B}}\right)}{A-B-A\,e^{2\,\sqrt{A-B}}+B\,e^{2\,\sqrt{A-B}}}
\end{align*}

 We shall show that in this set up partial uphill diffusion is possible. To this aim, because of the great number of parameters we specialize (\ref{solutionODE}) to a particular choice, namely
	\begin{equation}\label{refParamUPHILL}
		\sigma_{11}=\sigma_{22}=\Upsilon=1\qquad
		\sigma_{21}=\sigma_{12}=\frac{1}{2}\,.
\end{equation}
%By the knowledge of the physical constants we explicitly write:
The stationary profiles become
\begin{equation}
	\begin{split}
		\rho^{(\zeta)}(x)=&\frac{\rho_{L}^{(1)}}{2}+\frac{\rho_{L}^{(2)}}{2}-\frac{x\,\left(\rho_{L}^{(1)}+\rho_{L}^{(2)}-\rho_{R}^{(1)}-\rho_{R}^{(2)}\right)}{2}\\+(-1)^{\zeta}&\frac{{\mathrm{e}}^{2-2\,x}\,\left(\rho_{R}^{(1)}-\rho_{R}^{(2)}-\rho_{L}^{(1)}\,{\mathrm{e}}^2+\rho_{L}^{(2)}\,{\mathrm{e}}^2\right)}{2\,\left({\mathrm{e}}^4-1\right)}+(-1)^{\zeta}\frac{{\mathrm{e}}^{2\,x}\,\left(\rho_{L}^{(1)}-\rho_{L}^{(2)}-\rho_{R}^{(1)}\,{\mathrm{e}}^2+\rho_{R}^{(2)}\,{\mathrm{e}}^2\right)}{2\,\left({\mathrm{e}}^4-1\right)}\qquad\forall \zeta=1,2
	\end{split}
\end{equation}
and the diffusive currents read
\begin{equation}\label{current1}
	\begin{split}
		J^{(\zeta)}(x)=&\frac{3\,\rho_{L}^{(1)}}{4}+\frac{3\,\rho_{L}^{(2)}}{4}-\frac{3\,\rho_{R}^{(1)}}{4}-\frac{3\,\rho_{R}^{(2)}}{4}\\&+(-1)^{\zeta}\frac{{\mathrm{e}}^{2-2\,x}\,\left(\rho_{R}^{(1)}-\rho_{R}^{(2)}-\rho_{L}^{(1)}\,{\mathrm{e}}^2+\rho_{L}^{(2)}\,{\mathrm{e}}^2\right)}{2\,\left({\mathrm{e}}^4-1\right)}-(-1)^{\zeta}\frac{{\mathrm{e}}^{2\,x}\,\left(\rho_{L}^{(1)}-\rho_{L}^{(2)}-\rho_{R}^{(1)}\,{\mathrm{e}}^2+\rho_{R}^{(2)}\,{\mathrm{e}}^2\right)}{2\,\left({\mathrm{e}}^4-1\right)}\qquad \forall\zeta= 1,2
	\end{split}
\end{equation}
The problem of having partial uphill for, say, the species 1 is then the following:
by assuming that $\rho_{L}^{(1)}<\rho_{R}^{(1)}$
\begin{equation}\label{LocalUphillProblemWeak}
	\text{find }\quad(\rho_{L}^{(1)},\rho_{L}^{(2)},\rho_{R}^{(1)},\rho_{R}^{(2)})\;\;\;\text{ such that }\;\;\; \min_{x\in [0,1]}J^{(1)}(x)>0.
\end{equation}
There are choices of boundary densities that allow for partial uphill diffusion of the species 1. We give an example in Figure \ref{fig:uno}. 

A similar analysis can be done for the discretized
equations \eqref{goalBdl}, \eqref{goalBulk}, \eqref{goalBdr}.

\section{A two-parameter family of models}\label{appendixa}
In the following we report the matrices that describe the two-parameter family of generators introduced in Remark \ref{remarkFamily}. The matrices representing the generators $\mathcal{L}_{z,z+1}$ are of dimension $9\times 9$ while the matrices representing the generators $\mathcal{L}_{1},\;\mathcal{L}_{N}$ are of dimension $3\times 3$. The elements of these matrices are ordered as follows: 
\begin{itemize}
    \item for $\mathcal{L}_{z,z+1}$, the row and the column indexes are $$00,01,02,10,11,12,20,21,22$$ For example, the element on the $3^{\text{rd}}$ row and $4^{\text{th}}$ column gives the rate of transition $02 \to 10$ 
    \item for the site matrices $\mathcal{L}_{1}$ and $\mathcal{L}_{N}$, the rows and the columns a indexes are $0,1,2$. 
\end{itemize}
\begin{equation}\label{leftGeneral}
\begin{split}
&\mathcal{L}_{1}=\\
&\scalebox{0.85}{$\begin{pmatrix}
		- \sigma_{11}\rho_{L}^{(1)} - \sigma_{12}\rho_{L}^{(2)} - \sigma_{21}\rho_{L}^{(1)} - \sigma_{22}\rho_{L}^{(2)}&\sigma_{11}\rho_{L}^{(1)}+\sigma_{12}\rho_{L}^{(2)}&\sigma_{21}\rho_{L}^{(1)}+\sigma_{22}\rho_{L}^{(2)}\\ \\
		\sigma_{11} + \sigma_{21} - \sigma_{11}\rho_{L}^{(1)} - \sigma_{12}\rho_{L}^{(2)} - \sigma_{21}\rho_{L}^{(1)} - \sigma_{22}\rho_{L}^{(2)}&\sigma_{11}\rho_{L}^{(1)} - \sigma_{21} - h - \sigma_{11} + \sigma_{12}\rho_{L}^{(2)}&h + \sigma_{21}\rho_{L}^{(1)} + \sigma_{22}\rho_{L}^{(2)}\\ \\
		\sigma_{22}+\sigma_{12}-\sigma_{22}\rho_{L}^{(2)}-\sigma_{21}\rho_{L}^{(1)}-\sigma_{12}\rho_{L}^{(2)}-\sigma_{11}\rho_{L}^{(1)}&m + \sigma_{11}\rho_{L}^{(1)} + \sigma_{12}\rho_{L}^{(2)}& \sigma_{21}\rho_{L}^{(1)} - \sigma_{12} - m - \sigma_{22} + \sigma_{22}\rho_{L}^{(2)}
	\end{pmatrix}$}
	\end{split}
\end{equation}

\begin{equation}\label{explicitGeneratoBulk}
	\begin{split}
		&\mathcal{L}_{z,z+1}=\\
		&\scalebox{0.85}{	$\begin{pmatrix}
				\Gamma_{00}^{00}&0&0&0&0&0&0&0&0\\ \\
				0&\Gamma_{01}^{01}&h&\sigma_{11}&0&0&\sigma_{21}&0&0\\ \\
				0&m&\Gamma_{02}^{02}&\sigma_{12}&0&0&\sigma_{22}&0&0\\ \\
				0&\sigma_{11}&\sigma_{21}&\Gamma_{10}^{10}&0&0&\Upsilon-2\sigma_{21}-h&0&0\\ \\
				0&0&0&0&\Gamma_{11}^{11}&h&0&\Upsilon-2\sigma_{21}-h&\sigma_{21}\\ \\
				0&0&0&0&m&\Gamma_{12}^{12}&0&\sigma_{11}&\Upsilon-\sigma_{12}-\sigma_{21}-h\\ \\
				0&\sigma_{12}&\sigma_{22}&\Upsilon-2\sigma_{12}-m&0&0&\Gamma_{20}^{20}&0&0\\ \\
				0&0&0&0&\Upsilon-\sigma_{12}-\sigma_{21}-m&\sigma_{22}&0&\Gamma_{21}^{21}&h\\ \\
				0&0&0&0&\sigma_{12}&\Upsilon-2\sigma_{12}-m&0&m&\Gamma_{22}^{22}
			\end{pmatrix}$}
	\end{split}
\end{equation}
Due to the stochasticity of the generator, the diagonal elements are the following
\begin{align*}
	&\Gamma_{00}^{00}=0& &\Gamma_{01}^{01}-\sigma_{11}-\sigma_{21}-h& &\Gamma_{02}^{02}=-\sigma_{22}-\sigma_{12}-m\\ &\Gamma_{10}^{10}=-\Upsilon-\sigma_{11}+\sigma_{21}+h&&\Gamma_{11}^{11}=-\Upsilon+\sigma_{21}& &
	\Gamma_{12}^{12}=-\sigma_{11}-\Upsilon+\sigma_{12}+\sigma_{21}-m+h\\ &\Gamma_{20}^{20}=-\Upsilon-\sigma_{22}+\sigma_{12}+m&  &\Gamma_{21}^{21}=-\Upsilon-\sigma_{22}+\sigma_{21}+\sigma_{12}+m-h& &\Gamma_{22}^{22}=-\Upsilon+\sigma_{12}&
\end{align*}
\begin{equation}\label{rightGeneral}
\begin{split}
&\mathcal{L}_{N}=\\
&\scalebox{0.85}{$\begin{pmatrix}
		- \sigma_{11}\rho_{R}^{(1)} - \sigma_{12}\rho_{R}^{(2)} - \sigma_{21}\rho_{R}^{(1)} - \sigma_{22}\rho_{R}^{(2)}&\sigma_{11}\rho_{R}^{(1)}+\sigma_{12}\rho_{R}^{(2)}&\sigma_{21}\rho_{R}^{(1)}+\sigma_{22}\rho_{R}^{(2)}\\ \\
		\sigma_{11} + \sigma_{21} - \sigma_{11}\rho_{R}^{(1)} - \sigma_{12}\rho_{R}^{(2)} - \sigma_{21}\rho_{R}^{(1)} - \sigma_{22}\rho_{R}^{(2)}&\sigma_{11}\rho_{R}^{(1)} - \sigma_{21} - h - \sigma_{11} + \sigma_{12}\rho_{R}^{(2)}&h + \sigma_{21}\rho_{R}^{(1)} + \sigma_{22}\rho_{R}^{(2)}\\ \\
		\sigma_{22}+\sigma_{12}-\sigma_{22}\rho_{R}^{(2)}-\sigma_{21}\rho_{R}^{(1)}-\sigma_{12}\rho_{R}^{(2)}-\sigma_{11}\rho_{R}^{(1)}&m + \sigma_{11}\rho_{R}^{(1)} + \sigma_{12}\rho_{R}^{(2)}& \sigma_{21}\rho_{R}^{(1)} - \sigma_{12} - m - \sigma_{22} + \sigma_{22}\rho_{R}^{(2)}
	\end{pmatrix}$}
	\end{split}
\end{equation}

\section{Details of the proof of Theorem \ref{THM_principal}}\label{appendice}
	\subsection{Bulk process}
To solve (\ref{LinearBIGsystem}) it is useful to rewrite the system by using the following variables, that are made by sums of three non diagonal rates:
	\begin{align*}
		&y_{1}=\sum_{\beta=0}^{2}\Gamma_{10}^{\beta 1}&
		&y_{2}=\sum_{\beta=0}^{2}\Gamma_{00}^{\beta 1}&
		&y_{3}=\sum_{\beta=0}^{2}\Gamma_{01}^{1 \beta}&
		&y_{4}=\sum_{\beta=0}^{2}\Gamma_{00}^{1\beta}&
		&y_{5}=\sum_{\beta=0}^{2}\Gamma_{10}^{0 \beta}&
		&y_{6}=\sum_{\beta=0}^{2}\Gamma_{10}^{2\beta}\\
		&y_{7}=\sum_{\beta=0}^{2}\Gamma_{01}^{\beta 0}&
		&y_{8}=\sum_{\beta=0}^{2}\Gamma_{01}^{\beta 2}&
		&y_{9}=\sum_{\beta=0}^{2}\Gamma_{20}^{\beta 1}&
		&y_{10}=\sum_{\beta=0}^{2}\Gamma_{02}^{1\beta }&
		&y_{11}=\sum_{\beta=0}^{2}\Gamma_{02}^{\beta 1}&
		&y_{12}=\sum_{\beta=0}^{2}\Gamma_{20}^{1\beta}\\
		&y_{13}=\sum_{\beta=0}^{2}\Gamma_{20}^{\beta 2}&
		&y_{14}=\sum_{\beta=0}^{2}\Gamma_{00}^{\beta 2}&
		&y_{15}=\sum_{\beta=0}^{2}\Gamma_{02}^{2\beta}&
		&y_{16}=\sum_{\beta=0}^{2}\Gamma_{00}^{2\beta}&
		&y_{17}=\sum_{\beta=0}^{2}\Gamma_{20}^{0\beta}&
		&y_{18}=\sum_{\beta=0}^{2}\Gamma_{02}^{\beta 0}\\
		&y_{19}=\sum_{\beta=0}^{2}\Gamma_{10}^{\beta 2}&
		&y_{20}=\sum_{\beta=0}^{2}\Gamma_{01}^{2\beta}&
		&y_{21}=\sum_{\beta=0}^{2}\Gamma_{11}^{\beta 0}&
		&y_{22}=\sum_{\beta=0}^{2}\Gamma_{21}^{\beta 0}&
		&y_{23}=\sum_{\beta=0}^{2}\Gamma_{22}^{\beta 1}&
		&y_{24}=\sum_{\beta=0}^{2}\Gamma_{11}^{0\beta}\\
		&y_{25}=\sum_{\beta=0}^{2}\Gamma_{12}^{0\beta}&
		&y_{26}=\sum_{\beta=0}^{2}\Gamma_{12}^{\beta 1}&
		&y_{27}=\sum_{\beta=0}^{2}\Gamma_{21}^{1\beta}&
		&y_{28}=\sum_{\beta=0}^{2}\Gamma_{22}^{1\beta}&
		&y_{29}=\sum_{\beta=0}^{2}\Gamma_{11}^{\beta 2}&
		&y_{30}=\sum_{\beta=0}^{2}\Gamma_{12}^{\beta 0}\\
		&y_{31}=\sum_{\beta=0}^{2}\Gamma_{21}^{\beta 2}&
		&y_{32}=\sum_{\beta=0}^{2}\Gamma_{22}^{\beta 0}&
		&y_{33}=\sum_{\beta=0}^{2}\Gamma_{11}^{2 \beta}&
		&y_{34}=\sum_{\beta=0}^{2}\Gamma_{12}^{2\beta}&
		&y_{35}=\sum_{\beta=0}^{2}\Gamma_{21}^{0 \beta}&
		&y_{36}=\sum_{\beta=0}^{2}\Gamma_{22}^{0\beta}
	\end{align*}
Let us introduce the following: 
\begin{itemize}
	\item \textit{unknown vector}:  $\mathbf{y}\in \mathbb{R}_+^{36}$
	\begin{equation*}
		\mathbf{y}=(y_{i})_{i=1,\ldots 36}
	\end{equation*}
	\item \textit{known term}:  $\mathbf{b}\in \mathbb{R}^{30}$  (that is exactly the one in (\ref{LinearBIGsystem}))
	\begin{equation*}
		\begin{split}
			\mathbf{b}=\left(\sigma_{11},\sigma_{11},-2\sigma_{11}-\Upsilon,\sigma_{12},\sigma_{12},-2\sigma_{12}+\Upsilon,\sigma_{22},\sigma_{22},-2\sigma_{22}-\Upsilon,\sigma_{21},\sigma_{21},-2\sigma_{21}+\Upsilon,\right.\\ \left.0,0,0,0,0,0,0,0,0,0,0,0,0,0,0,0,0,0\right)^{T}
		\end{split}
	\end{equation*}
	\item \textit{coefficient matrix}:  $\Xi\in\mathbb{R}^{30\times 36}$ (that is full rank) 
\end{itemize}
By using the above vectors and matrix, the system (\ref{LinearBIGsystem}) can be rewritten as
\begin{equation}\label{semplificato}
	\Xi \mathbf{y}=\mathbf{b}.
\end{equation}
The systems (\ref{LinearBIGsystem}) and (\ref{semplificato}) are two ways of writing the conditions (\ref{closureConditions}), (\ref{Laplacianconditions}), (\ref{annulmentKnownTerm}).
By consequence,
there exists an other full rank matrix, say $\Lambda\in \mathbb{R}^{36\times 72}$, that allows to retrieve a $36$ parameter family of solutions of (\ref{LinearBIGsystem}) once we know the one of (\ref{semplificato}) as follows
\begin{equation}\label{intermediate}
		\Lambda \mathbf{u}=\mathbf{y}.
	\end{equation}
	We first solve \eqref{semplificato} and then we retrieve the specific solution \eqref{explicitGeneratoBulk} of \eqref{LinearBIGsystem}, by solving  \eqref{intermediate} with some specific choices of the $36$ parameters.\\
	%The advantage of this change of variables is the reduction of the under-determination order from $36$ to $6$ only. This simplification allows to find explicitly a six parameter family of non negative solutions of \eqref{semplificato}, and retrieve the one of  \eqref{LinearBIGsystem} by making assumptions that allow to remove the under-determination. Thus, we split the proof in two steps.  
%Let us observe that the system (\ref{intermediate}) is underdetermined and thus has infinite many solutions.
\newline
\indent \textbf{Solution of (\ref{semplificato}}): the under-determination order is $6$ and thus 6 components of the vector $\mathbf{y}$ are, actually, free parameters. Without any constraint \eqref{semplificato} would have a 6 parameter family of solutions. However, the non-negativity of the solution (the $y_{i}$ are sums of transition rates) will reduce the dependence on just two free parameters. \\
Indeed, by direct computations and by recalling that the variables $\{y_{j}\}_{j=1,\ldots 36}$  must be non-negative we find the following $12$ unknowns by using just $10$ equations, namely:
\begin{align*}
   & y_{1}-y_{2}=\sigma_{11}& &y_{3}-y_{4}=\sigma_{11}&&y_{9}-y_{2}=\sigma_{12}&&y_{10}-y_{4}=\sigma_{12}&&y_{13}-y_{14}=\sigma_{22}\\
   &y_{15}-y_{16}=\sigma_{22}&&y_{19}-y_{14}=\sigma_{21}&&y_{20}-y_{16}=\sigma_{21}&&y_{2}+y_{14}=0&&y_{4}+y_{16}=0
\end{align*}
that are solved if and only if
\begin{align*}
		&y_{2}=y_{4}=y_{14}=y_{16}=0&
		&y_{1}=y_{3}=\sigma_{11}		&y_{19}=y_{20}=\sigma_{21}\\
		&y_{9}=y_{10}=\sigma_{12}&
		&y_{13}=y_{15}=\sigma_{22}.
\end{align*}
By the non negativeness of the above $y_{j}$, it follows that
\begin{equation}\label{primeCondizioni}
	\sigma_{11},\sigma_{12},\sigma_{21},\sigma_{22}\geq 0.
\end{equation}
Now, it remains to solve a system with $20$ equations and $24$ unknowns. By introducing as parameters $(y_{7},y_{8},y_{11},y_{17}):=(g,h,m,s)$, this $20\times 24$ system becomes a $20\times 20$ parametric system.  This last one has the following explicit parametric solution: 
%This means that two (on six) parameters are fixed at zero, as a consequence the under-determination order is now just $4$. 

%After this first computation, we are left with a system of $20$ equations in $24$ unknowns. Since the matrix $\Xi$ is full rank, we rewrite this system in a $20\times 20$ form by introducing four parameters, that are renamed as:
%\begin{equation}
%	(y_{7},y_{8},y_{11},y_{17})=(g,h,m,s)
%\end{equation}
%The parametric solution of this $20\times 20$ system is given by:
\begin{equation}\label{soluzioneVenti}
\begin{split}
	&\left(
		y_{5},\;
		y_{6},\;
		y_{12},\;
		y_{18},\;
		y_{21},\;
		y_{22},\;
		y_{23},\;
		y_{24},\;
		y_{25},\;
		y_{26},\;
		y_{27},\;
		y_{28},\;
		y_{29},\;
		y_{30},\;
		y_{31},\;
		y_{32},\;
		y_{33},\;
		y_{34},\;
		y_{35},\;
		y_{36}
\right)=\\ 
		&\left(2\sigma_{11} + 2\sigma_{21} - g,\;
		\Upsilon - 2\sigma_{21} - h,\;
		\Upsilon - 2\sigma_{12} - m,\;
		2\sigma_{12} + 2\sigma_{22} - s,\;
		g - \sigma_{21} - \sigma_{11},\;
		g - \sigma_{22} - \sigma_{12},\;
		\sigma_{12} + m,\right.\\& \left.
		\sigma_{11} + \sigma_{21} - g,
		2\sigma_{11} - \sigma_{12} + 2\sigma_{21} - \sigma_{22} - g,\;
		\sigma_{11} + m,\;
		\sigma_{11} - 2\sigma_{12} + \sigma - m,
		\Upsilon- \sigma_{12} - m,\;
		\sigma_{21} + h,\right. \\ &\left.
		2\sigma_{12} - \sigma_{11} - \sigma_{21} + 2\sigma_{22} - s,\;
		\sigma_{22} + h,  \;
		\sigma_{12} + \sigma_{22} - s,\;
		\Upsilon - \sigma_{21} - h,\;
		\sigma_{22} - 2\sigma_{21} + \Upsilon - h,
		s - \sigma_{21} - \sigma_{11},\right. \\ & \left.
		s - \sigma_{22} - \sigma_{12}\right).
	\end{split}
\end{equation}
Since all the $y_{i}$ are sums of non negative transition rates, we impose that the components of \eqref{soluzioneVenti} are non negative. This is true if and only if:
\begin{equation}\label{ViaParametri}
		s=\sigma_{11}+\sigma_{21}\quad
		g=\sigma_{11}+\sigma_{21}
\end{equation}
and 
\begin{equation}\label{secondeCondizioni}
	\Upsilon,h,m\geq 0\qquad
	\sigma_{12}\leq \frac{\Upsilon-m}{2}\qquad
	\sigma_{21}\leq \frac{\Upsilon-h}{2}
\qquad \sigma_{11}+\sigma_{21}=\sigma_{12}+\sigma_{22}.\end{equation}
Since (\ref{ViaParametri}) fixes the value of two of the four parameters, the non negative solution only depends on $h,m$. Putting together (\ref{primeCondizioni}) and (\ref{secondeCondizioni}) we obtain (\ref{conditionTHM}). Finally, this explicit non-negative solution of (\ref{semplificato}) is 
\begin{equation}\label{generalSolution}
\begin{split}
	\mathbf{y}=
\left(  \sigma_{11},\;	
0	,\;
\sigma_{11}\;
0	,\;
\sigma_{11} + \sigma_{21}	,\;
\Upsilon - 2\sigma_{21} - h	,\; 
\sigma_{11} + \sigma_{21}	,\;
h	,\;
\sigma_{12}	,\;
\sigma_{12}	,\;
m	,\;
\Upsilon - 2\sigma_{12} - m,\right.\\ \left.
\sigma_{11} - \sigma_{12} + \sigma_{21}	,\;
0	,\;
\sigma_{11} - \sigma_{12} + \sigma_{21}	,\;
0	,\;
\sigma_{11} + \sigma_{21}	,\;
\sigma_{11} + \sigma_{21}
\sigma_{21}	,\;
\sigma_{21}	,\;
0	,\;
0	,\;
\sigma_{12} + m	,\;
0,\;
0	,\;
\sigma_{11} + m	,\right.	\\ \left.
\sigma_{11} - 2\sigma_{12} + \Upsilon - m,\;	
\Upsilon - \sigma_{12} - m	,\;
\sigma_{21} + h,\;
0,\;
\sigma_{11} - \sigma_{12} + \sigma_{21} + h	,\;
0	,\;
\Upsilon - \sigma_{21} - h	,\right.	\\ \left.
\sigma_{11} - \sigma_{12} - \sigma_{21} + \Upsilon - h	,\;
0	,\;
0\right)
\end{split}
\end{equation}
\newline
 \indent \textbf{Solution of (\ref{LinearBIGsystem})}: from \eqref{generalSolution} we know the explicit solution of \eqref{semplificato}. To find the solution of \eqref{LinearBIGsystem}, we solve \eqref{intermediate}. This last system is full rank. It has $72$ unknowns in $36$ equations, thus the order of under-determination is $36$.  We must look for non-negative solution. To remove the under-determination, and produce examples \eqref{explicitGeneratoBulk} we impose the following conditions: 
\begin{enumerate}[i]
	\item The matrix associated to the generator has the greater number of zeros;
	\item Fix the following rates:  \begin{equation}
		\Gamma_{12}^{21}=\sigma_{11}\quad 	\Gamma_{21}^{12}=\sigma_{22}\quad
	\Gamma_{11}^{22}=\sigma_{21}\quad
	\Gamma_{22}^{11}=\sigma_{12}.
	\end{equation}
\end{enumerate}

With the above two requests, the solution of (\ref{intermediate}) is unique (for fixed parameters $h,m$ and for fixed diffusivity matrix and reaction constant) and the bulk generator takes the form (\ref{explicitGeneratoBulk}). Indeed, by considering (\ref{generalSolution}) we have:
\begin{itemize} 
	\item The row $\Gamma_{00}^{\alpha,\beta}$ has all the elements are zero;
	\item The row $\Gamma_{01}^{\alpha,\beta}$ is found by solving 
	\begin{align*}&\Gamma_{01}^{10}+\Gamma_{01}^{11}+\Gamma_{01}^{12}=\sigma_{11}&
	 &\Gamma_{01}^{00}+\Gamma_{01}^{10}+\Gamma_{01}^{20}=\sigma_{11}+\sigma_{21}\\
		 &\Gamma_{01}^{02}+\Gamma_{01}^{12}+\Gamma_{01}^{22}=h&
		 &\Gamma_{01}^{20}+\Gamma_{01}^{21}+\Gamma_{01}^{22}=\sigma_{21}.
	\end{align*}
By the conditions $i$ and $ii$ previously required, we obtain $\Gamma_{01}^{10}=\sigma_{11}$, $\Gamma_{01}^{20}=\sigma_{12}$, $\Gamma_{01}^{02}=h$ and all the other off-diagonal rates are equal to zero. By similar arguments, also the rows $\Gamma_{02}^{\alpha\beta}, \Gamma_{10}^{\alpha\beta}, \Gamma_{20}^{\alpha\beta}$ are determined.
\item {The row $\Gamma_{11}^{\alpha\beta}$} is found by solving:
\begin{align*}
		&\Gamma_{11}^{02}+\Gamma_{11}^{12}+\Gamma_{11}^{22}=\sigma_{21}+h&
		&\Gamma_{11}^{20}+\Gamma_{11}^{21}+\Gamma_{11}^{22}=\Upsilon-\sigma_{21}-h\\
		&\Gamma_{11}^{00}+\Gamma_{11}^{10}+\Gamma_{11}^{20}=0&
		&\Gamma_{11}^{00}+\Gamma_{11}^{01}+\Gamma_{11}^{02}=0.
\end{align*}
By the conditions $i$ and $ii$ previously required we obtain $\Gamma_{11}^{22}=\sigma_{21}$, $\Gamma_{11}^{12}=h$, $\Gamma_{11}^{21}=\Upsilon-2\sigma_{21}-h$ and all the other off-diagonal rates are equal to zero. By similar arguments, also the rows  $\Gamma_{12}^{\alpha\beta},\Gamma_{21}^{\alpha\beta},\Gamma_{22}^{\alpha\beta}$ are determined.
	\end{itemize}
We observe that, when $h=m=0$	\eqref{explicitGeneratoBulk} do coincide with the non negative least square solution (see \cite{boullion1971generalized}) of \eqref{intermediate}. 
	 \eqref{example1par} is recovered from \eqref{explicitGeneratoBulk} when $\sigma_{21}=\sigma_{12}$,  $\sigma_{22}=\sigma_{11}$ and $h=m$ in \eqref{explicitGeneratoBulk}.
\subsection{Boundary processes}
Once the bulk is known, the conditions for the boundaries form two determined systems of linear algebraic equations. We solve explicitly only the left boundary; the solution of the right one is very similar.  \\ \newline
\indent\textbf{Left boundary}: recalling the definitions of $B_{1}$ and $C_{2}$, we have the following
\begin{align*}
		&B_{1}^{11}=-y_{5}-y_{6}-y_{4}&
		&B_{1}^{12}=y_{12}-y_{4}&
		&B_{1}^{21}=y_{6}-y_{16}&
		&B_{1}^{22}=-y_{17}-y_{12}-y_{16}\\
	&C_{2}^{11}=-y_{7}-h-y_{2}&
	&C_{2}^{12}=m-y_{2}&
	&C_{2}^{21}=h-y_{14}&
		&C_{2}^{22}=-y_{18}-m-y_{14};
\end{align*}
by consequence system (\ref{BoundarySystemL}) is rewritten as:
\begin{equation*}
	\begin{pmatrix}
		1	&	0	&	0	&	0	&	0	&	0	\\
		-1	&	0	&	-1	&	-1	&	0	&	0	\\
		-1	&	0	&	0	&	0	&	0	&	1	\\
		0	&	1	&	0	&	0	&	0	&	0	\\
		0	&	-1	&	0	&	1	&	0	&	0	\\
		0	&	-1	&	0	&	0	&	-1	&	-1	
	\end{pmatrix}\begin{pmatrix}
		W_{0}^{1}(1)\\
		W_{0}^{2}(1)\\
		W_{1}^{0}(1)\\
		W_{1}^{2}(1)\\
		W_{2}^{0}(1)\\
		W_{2}^{1}(1)
	\end{pmatrix}=\begin{pmatrix}
		\sigma_{11}\rho_{L}^{(1)}+\sigma_{12}\rho_{L}^{(2)}\\
		-\sigma_{11}-\sigma_{21}-h\\
		m\\
		\sigma_{21}\rho_{L}^{(1)}+\sigma_{22}\rho_{L}^{(2)}\\
		h\\
		-\sigma_{22}-\sigma_{12}-m
	\end{pmatrix}.
\end{equation*}
The coefficient matrix of the above system has full rank; thus there exists a unique solution. Recalling the definition of $W_{\gamma}^{\alpha}(1)$ we obtain \eqref{leftGeneral}. As a consequence of \eqref{conditionTHM}, and in particular $\sigma_{11}+\sigma_{21}=\sigma_{12}+\sigma_{22}$, this generator has non negative non-diagonal transition rates if
\begin{equation}\label{constraintBC1}
	0\leq \rho_{L}^{(1)}+\rho_{L}^{(2)}\leq 1.
\end{equation}
\eqref{constraintBC1} is always true since we assumed that since we assumed that the sum of the densities of the two species in the reservoir is at most one.\\ \newline
\indent \textbf{Right boundary}: by  similar arguments we solve \eqref{BoundarySystemR} and we obtain the right boundary, i.e. \eqref{rightGeneral}. This matrix has non-negative off-diagonal rates if: 
\begin{comment}
\begin{equation*}
	\mathcal{L}_{N}=
	\scalebox{0.73}{$\begin{pmatrix}
			- \sigma_{11}\rho_{R}^{(1)} - \sigma_{12}\rho_{R}^{(2)} - \sigma_{21}\rho_{R}^{(1)} - \sigma_{22}\rho_{R}^{(2)}&\sigma_{11}\rho_{R}^{(1)}+\sigma_{12}\rho_{R}^{(2)}&\sigma_{21}\rho_{R}^{(1)}+\sigma_{22}\rho_{R}^{(2)}\\
			\sigma_{11} + \sigma_{21} - \sigma_{11}\rho_{R}^{(1)} - \sigma_{12}\rho_{R}^{(2)} - \sigma_{21}\rho_{R}^{(1)} - \sigma_{22}\rho_{R}^{(2)}&\sigma_{11}\rho_{R}^{(1)} - \sigma_{21} - h - \sigma_{11} + \sigma_{12}\rho_{R}^{(2)}&h + \sigma_{21}\rho_{R}^{(1)} + \sigma_{22}\rho_{R}^{(2)}\\
			\sigma_{22}+\sigma_{12}-\sigma_{22}\rho_{R}^{(2)}-\sigma_{21}\rho_{R}^{(1)}-\sigma_{12}\rho_{R}^{(2)}-\sigma_{11}\rho_{R}^{(1)}&m + \sigma_{11}\rho_{R}^{(1)} + \sigma_{12}\rho_{R}^{(2)}& \sigma_{21}\rho_{R}^{(1)} - \sigma_{21} - m - \sigma_{11} + \sigma_{22}\rho_{R}^{(2)}
		\end{pmatrix}$}
\end{equation*}
and thus we must require that:.
\end{comment}
\begin{equation}\label{constraintBCN}
	0\leq \rho_{R}^{(1)}+\rho_{R}^{(2)}\leq 1.
\end{equation}
\eqref{constraintBCN} is always true since we assumed that the sum of the densities in the reservoir is at most one.
\end{appendices}
\bibliographystyle{unsrt}
\bibliography{biblio}
\end{document}